\documentclass[10pt,aps,prl,twocolumn,notitlepage,nofootinbib,nopacs,final,letterpaper,superscriptaddress,longbibliography,floatfix]{revtex4-2}

\usepackage[utf8]{inputenc}
\usepackage{amsmath,amssymb,amsthm}
\usepackage{array}
\usepackage{braket}
\usepackage{calc}
\usepackage{color}
\usepackage{dcolumn}
\usepackage[inline]{enumitem}
\usepackage{graphicx}
\usepackage[colorlinks=true,linkcolor=black,urlcolor=blue,citecolor=blue,anchorcolor=blue]{hyperref}
\usepackage{latexsym}
\usepackage{multirow}
\usepackage{mathtools}
\usepackage{subfigure}
\usepackage{txfonts}
\usepackage{braket}
\DeclareGraphicsExtensions{.jpg,.pdf, .mps, .png, .eps, .ps, .EPS,.gif}

\usepackage{pdfpages}
\makeatletter
\AtBeginDocument{\let\LS@rot\@undefined}
\makeatother

\begin{document}

\newcommand{\avg}[1]{\langle{#1}\rangle}
\newcommand{\Avg}[1]{\left\langle{#1}\right\rangle}

\newcommand{\gin}[1]{\textbf{\color{green}#1}}

\author{Guillaume St-Onge}
\email{guillaume.st-onge.4@ulaval.ca}
\affiliation{D\'epartement de physique, de g\'enie physique et d'optique, Universit\'e Laval, Qu\'ebec (Qu\'ebec), Canada G1V 0A6}
\affiliation{Centre interdisciplinaire en mod\'elisation math\'ematique, Universit\'e Laval, Qu\'ebec (Qu\'ebec), Canada G1V 0A6}
\author{Hanlin Sun}
\affiliation{School of Mathematical Sciences, Queen Mary University of London, London, E1 4NS, United Kingdom}
\author{Antoine Allard}
\affiliation{D\'epartement de physique, de g\'enie physique et d'optique, Universit\'e Laval, Qu\'ebec (Qu\'ebec), Canada G1V 0A6}
\affiliation{Centre interdisciplinaire en mod\'elisation math\'ematique, Universit\'e Laval, Qu\'ebec (Qu\'ebec), Canada G1V 0A6}
\affiliation{Vermont Complex Systems Center, University of Vermont, Burlington, VT 05405}
\author{Laurent H\'{e}bert-Dufresne}
\affiliation{D\'epartement de physique, de g\'enie physique et d'optique, Universit\'e Laval, Qu\'ebec (Qu\'ebec), Canada G1V 0A6}
\affiliation{Vermont Complex Systems Center, University of Vermont, Burlington, VT 05405}
\affiliation{Department of Computer Science, University of Vermont, Burlington, VT 05405}
\author{Ginestra Bianconi}
\affiliation{School of Mathematical Sciences, Queen Mary University of London, London, E1 4NS, United Kingdom}
\affiliation{The Alan Turing Institute, 96 Euston Rd, London NW1 2DB, United Kingdom}

\title{Universal nonlinear infection kernel from heterogeneous exposure on higher-order networks}

\begin{abstract}
The colocation of individuals in different environments is an important prerequisite for exposure to infectious diseases on a social network.
Standard epidemic models fail to capture the potential complexity of this scenario by (1) neglecting the higher-order structure of contacts which typically occur through environments like workplaces, restaurants, and households; and by (2) assuming a linear relationship between the exposure to infected contacts and the risk of infection.
Here, we leverage a hypergraph model to embrace the heterogeneity of environments and the heterogeneity of individual participation in these environments.
We find that combining heterogeneous exposure with the concept of minimal infective dose induces a universal nonlinear relationship between infected contacts and infection risk.
Under nonlinear infection kernels, conventional epidemic wisdom breaks down with the emergence of discontinuous transitions, super-exponential spread, and hysteresis.
\end{abstract}

\maketitle

Mathematical models of epidemics play an increasingly important role in public health efforts and pandemic preparedness~\cite{rivers2019using}.
By providing insights on the interplay of the biological and sociological aspects of epidemics, models can test potential interventions in silico and suggest potential outcomes~\cite{epstein2008why}.
However, large-scale forecasting comparisons show that statistical models often outperform mechanistic models that make assumptions about spreading dynamics \cite{biggerstaff2018results}.

In this letter, we look at the interplay of two commonly used assumptions in disease models: Random mixing and the linearity between infection risk and exposures to infected individuals. In almost all disease models, doubling the number of contacts between susceptible and infectious individuals doubles the risk of infection for the susceptible individuals.
Some past work in mathematical biology has considered nonlinear infection rates~\cite{liu1986influence,hethcote1991some}, but these models are rarely used in practice.
Other fields such as sociology do often consider generalized contagion models, often dubbed complex contagions \cite{centola2018how,Lehmann2018}.
In these complex contagions, having a nonlinear relationship between infection rate and sources of infection allows the model to consider mechanisms such as social reinforcement~\cite{centola2010spread}, where a set of multiple exposures can have more impact than the mere sum of unique exposures.

The mathematical convenience of assuming random mixing when modeling infectious diseases comes at the price that all contacts between susceptible and infectious individuals are effectively equivalent.
This assumption has often been lifted using heterogeneous mathematical models where individuals are distinguished by some individual features such as their intrinsic susceptibility or reaction to the infection \cite{busenberg1991global, feng2000endemic}, relaxing the mass-action assumption directly \cite{hethcote1987epidemiological,hethcote1996modeling}, or by specifying an underlying contact network~\cite{pastor-satorras2001epidemic,pastor-satorras2015epidemic}.

\begin{figure*}
\centering
\includegraphics[width=\linewidth]{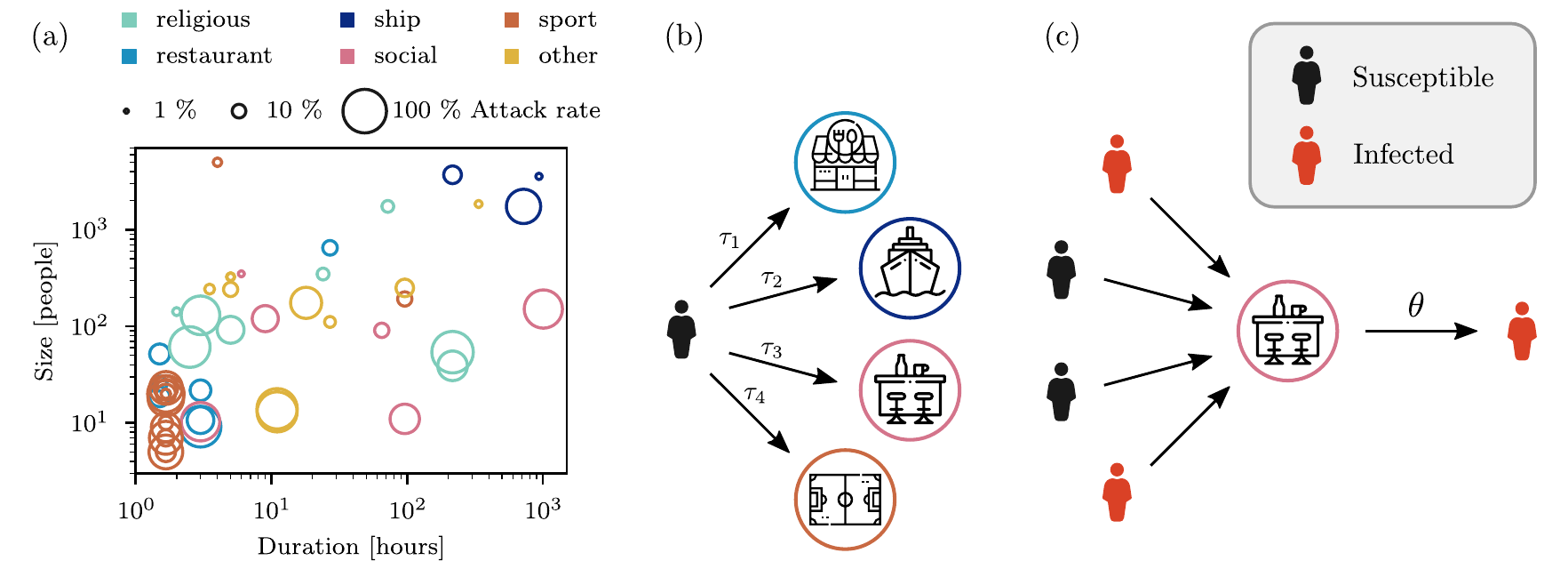}
\vspace{-2\baselineskip}
\caption{Modeling contagions and superspreading events through higher-order networks. (a) Scatter plot of superspreading events of COVID-19 where the number of people involved (size), the duration of the event, and the resulting proportion of infected individuals (attack rate) are all available (extracted from Refs.~\cite{SSdata_original,leclerc2020}, see the supplemental material \cite{SM}).
(b)-(c) Framework for contagions on hypergraphs~\cite{freepik}, where the size $m$ of the hyperedges (environment), the hyperdegree $k$ of the nodes (individuals), and the participation time to the environment $\tau$ are all heterogeneous, distributed according to $\hat{P}(m)$, $\tilde{P}(k)$, and $P(\tau)$ respectively. For the sake of simplicity, we assume the same distribution $P(\tau)$ for all environments. (b) At each time step $t$, an individual participates for a time $\tau$ (drawn independently) to each environment. (c) An individual gets infected with probability $\theta_m(\rho)$ in the environment at time step $t$, which depends on the size $m$ and the fraction infected $\rho$.}

\label{fig:sses}
\end{figure*}

Moreover, the linearity assumption says that all increments in the total exposure to infectious individuals (measured for example as a viral inoculum) are equivalent.
Evidence associated with the minimal infective dose of different infectious diseases shows that not all exposures are equal, and that some minimal dose might be required for an infection to likely occur.
More precisely, the ID$_{50}$ value is a measure of the dose needed to cause an infection in 50\% of individuals.
These concepts are needed because our immune system is usually able to handle microscopic challenges from viruses and bacteria alike.
While an infective dose of tuberculosis might only require between 1 and 5 bacteria~\cite{balasubramanian1994pathogenesis}, some enterics might require up to $10^9$ pathogenic particles~\cite{larocque2015syndromes}, and others like common respiratory infections still require further study~\cite{weber2008inactivation}.
There are indeed multiple different physical mechanisms behind immune evasion, for example some airborne viruses need to find their receptors on lung epithelial cells, while some bacteria might instead require interaction with the immune system \cite{gama2012immune}.
These mechanisms are reviewed in Refs. \cite{finlay1997common, hornef2002bacterial, lipsitch2007patterns, casadevall2008evolution, gama2012immune}, and all of them combine to determine the ID50 of specific pathogens.
Likewise, the decay or clearing rates of pathogens in non-infectious courses can also vary a lot, potentially requiring days for bacteria to hours for airborne viral infections.
For example, mathematical models for the pathogenesis of \mbox{SARS-CoV-2} or influenza A use decay rates of the order of 7-18 hours but empirical estimates vary wildly (see Ref. \cite{du2020mathematical} and \cite{beauchemin2011review} and references therein).

To study the effect of simultaneously relaxing these two assumptions, we consider a social structure where individuals attend a certain number of environments such as work places, gyms, or supermarkets.
This division of contact structure in environments is motivated by the known role of superspreading events, which are for example critical to the ongoing spread of COVID-19 \cite{liu2020secondary,wong2020evidence,althouse2020superspreading,endo2020estimating,bi2020epidemiology,miller2020full,lau2020charac,nielsen2021covid}.
While variations at the individual level is often used to explain superspreading \cite{lloyd2005superspreading}, we focus here on the variability of environments and of temporal patterns \cite{karsai2018bursty,karsai2012universal,holme2012temporal,cattuto2010dynamics,stehle2010dynamical,zhao2011social,cencetti2020temporala} at the \textit{group} level, which undoubtedly affect epidemics \cite{althouse2020superspreading}, especially when a certain exposure within a certain time window is needed to confidently spark an infection.
Interestingly, available case data highlight how there is no expected size or duration for such events. Transmission is highly context dependent on the settings (e.g., ventilation) and activity (e.g., singing or shouting) such that the resulting superspreading events are heterogeneous in size, duration and attack rate, as shown in Fig.~\ref{fig:sses}(a).
Higher-order contact structures and heterogeneous temporal patterns are therefore key ingredients for more realistic models of spreading dynamics.

Mathematically, we represent the contact structure as a hypergraph \cite{berge1989hypergraphs,battiston2020networks,torres2020why} where each environment is described by a hyperedge connecting $m$ nodes (individuals) and where each node is incident to $k$ hyperedges.
All hyperedges of a same size $m$ are considered equivalent, although this assumption is relaxed in the supplemental material to consider additional sources of heterogeneity \cite{SM}.
To model heterogeneous temporal patterns, we consider a discrete-time process, where at each time step $t = 1, 2, \dots$, we draw for each individual a \textit{participation time} $\tau \in [1,\tau_\mathrm{max}]$ for each environment to which they are connected [Fig~\ref{fig:sses}(b)].
The time steps correspond to fixed temporal windows of size $\tau_\mathrm{max}$, during which susceptible individuals can get infected through their participation to environments.

We first study the impact of the spatiotemporal co-location patterns on the \textit{infection kernel} $\theta_m(\rho)$, the probability of getting infected in an environment of size $m$ when a fraction $\rho$ of the other participants are infectious [Fig.~\ref{fig:sses}(c)].
We then analyze the properties of the resulting contagion process.

\paragraph{Universal infection kernel from heterogeneous exposure.}
Let us consider a susceptible individual participating to an environment of size $m$ for a duration $\tau$, where a fraction $\rho$ of the other participants are infectious.
During this exposure period, some of the other $m-1$ individuals might participate to the environment as well. We assume that the considered individual receives an infective dose $\kappa \in [0,\infty)$ from the infectious individuals, distributed according to $\pi(\kappa; \lambda)$, where $\lambda \equiv \langle \kappa \rangle$.
A reasonable assumption is that the mean dose received is proportional to the time spent in the environment and to the proportion of infectious people, $\lambda = \beta f(m) \tau \rho$, where $\beta$ is a rate of dose accumulation and $f(m)$, unitless, modulates the typical number of contacts in environments frequented by $m$ individuals.

While this is not a strict requirement for our results to hold (see supplemental material \cite{SM}), we further assume that the random variable for the dose can be written as $\kappa = \lambda u$, where $u$ is a random variable that is independent of $\lambda$.
In this case, $u$ can be seen as an intrinsic property of the contagion process---determined by rates of viral shedding, diffusion in the environment, variability of human interactions, etc.---while $\lambda$ acts as a \textit{scale} parameter, i.e., $\pi(\kappa; \lambda) = \pi(\kappa / \lambda;1)/\lambda \equiv \pi(\kappa / \lambda)/\lambda$.

To incorporate the concept of minimal infective dose, we assume that an individual develops the disease if $\kappa > K$, a perspective analogous to standard threshold models \cite{granovetter1978threshold,watts2002simple,dodds2004universal} and related to the assumption that successful host invasion necessitates multiple attempts by the pathogen \cite{anttila2017mechanistic}.
The probability of getting infected in the environment is then
\begin{align}
    \label{eq:ccdf_dose}
    \bar{\Pi}(K / \lambda) = \int_{K/\lambda}^\infty \pi(\kappa) \mathrm{d} \kappa\;.
\end{align}

\begin{figure*}
\centering
\includegraphics[width=\textwidth]{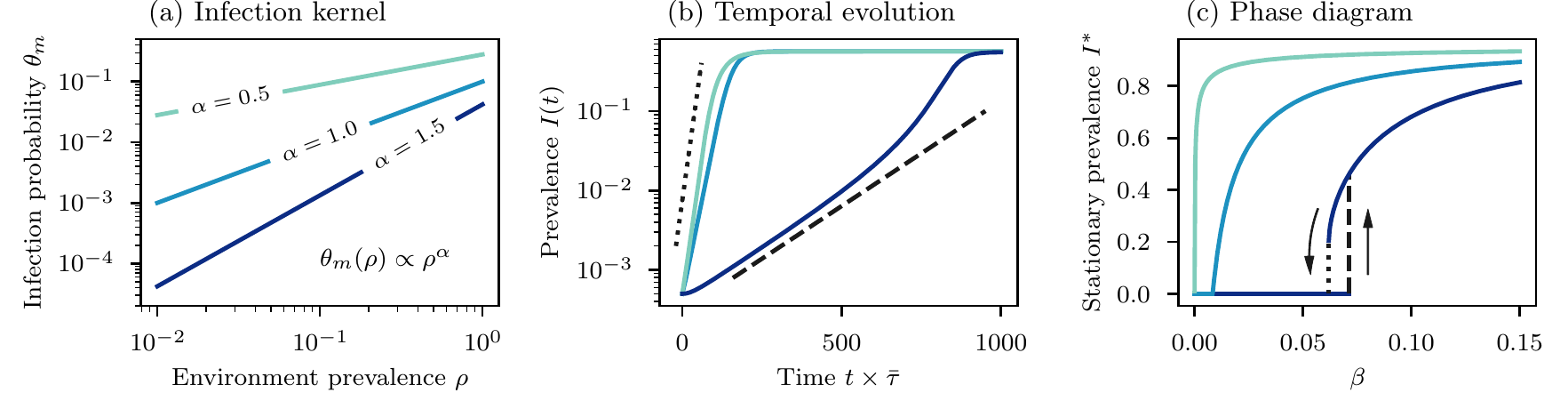}
\vspace{-2\baselineskip}
\caption{Properties of contagions with nonlinear infection kernels induced by heterogenous exposure. We use an exponential dose distribution $\pi(\kappa ; \lambda) \propto e^{-\kappa/\lambda}$ with a power-law distribution of participation time $P(\tau) \propto \tau^{-\alpha-1}$, a clearing window $\mathcal{T} \to \infty$, and $f(m) = 1$. (a) Effective infection kernel using $\beta = 0.1$. The infection probability has a power law scaling $\theta_m(\rho) \propto \rho^\alpha$. (b)-(c) We use Poisson distributions for both $\tilde{P}(k)$ and $\hat{P}(m)$, with $\langle k \rangle = 5$ and $\langle m \rangle = 10$, and set $\mu = 0.05$. We use Eqs.~(\ref{eq:rhok_time_evo}-\ref{eq:thetam_bar_def}) to evolve the system. (b)  Supra-linear kernels $\nu > 1$ lead to a super-exponential growth for the global prevalence $I(t)$. We use $\beta = 5\times 10^{-4}$, $\beta = 0.025$ and $\beta = 0.077$ for $\nu = 0.5$, $\nu = 1$ and $\nu = 1.5$ respectively. $\bar{\tau}$ is the median exposure period.
(c) The phase diagram for stable solutions in the stationary state $(t \to \infty)$ can be continuous or discontinuous with a bistable regime. Sub-linear and linear kernels $\nu \leq 1$ lead to a continuous phase transition, and the invasion threshold $\beta_\mathrm{c}$ vanishes for $\nu \to 0$. Supra-linear kernels $\nu > 1$ can lead to a discontinuous phase transition with a bistable regime.}
\label{fig:kernel}
\end{figure*}

The infection kernel $\theta_m(\rho)$ is calculated by averaging $\bar{\Pi}(K / \lambda)$ over $P(\tau)$.
We focus here on the case of heterogeneous exposure periods modeled with a Pareto distribution \mbox{$P(\tau)=C_{\alpha}\tau^{-\alpha-1}$}, where $C_{\alpha}$ is a normalization constant, $\alpha > 0$ and $\tau \in [1,\tau_\mathrm{max}]$.
However, for our dose mechanism to be well defined, we can only average over participation times $\tau \in [1,\mathcal{T}]$, where $\mathcal{T} \leq \tau_\mathrm{max}$ is the \textit{clearing window}, i.e., the characteristic time for the immune system to get rid of any dose $\kappa < K$.
If we assume that this clearing window is sufficiently large compared to $\tau_\mathrm{c} \equiv K / \beta f(m) \rho$, the characteristic time to get infected in the environment, we can neglect events where $\tau \geq \mathcal{T}$ as they do not contribute significantly to the infection kernel \cite{SM}.
We therefore redefine our support as $\tau \in [1,\mathcal{T}]$ and the infection kernel is
\begin{align}
    \label{eq:thetam_def}
    \theta_m(\rho)
      & = \int_1^{\mathcal{T}} \bar{\Pi}(\tau_\mathrm{c}/ \tau) P(\tau) \mathrm{d} \tau \\
      & = \frac{C_\alpha}{\alpha}\left[ \bar{\Pi}(\tau_\mathrm{c}) - \bar{\Pi}(\tau_\mathrm{c}/\mathcal{T})\mathcal{T}^{-\alpha} + \tau_\mathrm{c}^{-\alpha}\int_{\tau_\mathrm{c}/\mathcal{T}}^{\tau_\mathrm{c}} \pi(y) y^{\alpha} \mathrm{d}y \right] \;. \nonumber
\end{align}
When \mbox{$1 \ll \tau_\mathrm{c} \ll \mathcal{T}$} and $\pi(y)$ decreases faster than $y^{-\alpha-1}$, then the integral on the right converges to a constant, the term in $\mathcal{T}^{-\alpha}$ can be neglected, and $\bar{\Pi}(\tau_\mathrm{c}) \ll \tau_\mathrm{c}^{-\alpha}$, which implies
\begin{align}
    \label{eq:kernel_asymptotic}
    \theta_m(\rho) \sim D_\alpha \tau_\mathrm{c}^{-\alpha} \propto \rho^\alpha \;,
\end{align}
where $D_\alpha$ is some constant.
This form of infection kernel is \textit{universal}, driven by temporal patterns, and does not depend on the value of $K$ (given $K > 0$) or on the particular form of $\pi$.
We illustrate it in Fig.~\ref{fig:kernel}(a) using an exponential for the dose distribution---other cases such as the Weibull and the Fréchet distribution are presented in the supplemental material \cite{SM}.

Let us stress that the condition \mbox{$1 \ll \tau_\mathrm{c} \ll \mathcal{T}$} is not restrictive.
On the time scale on which we measure the exposure periods, $\tau_\mathrm{c} \gg 1$ implies that the contagion does not spread too fast, otherwise the whole population would be rapidly infected, while $\mathcal{T} \gg \tau_\mathrm{c}$ suggests that it is transmissible before the immune system is able to clear the dose received.

More broadly, we do not need to assume that $\pi(y)$ decreases faster than $y^{-\alpha-1}$, nor that $\lambda$ is a scale parameter for $\kappa$. In fact, $\pi(\kappa;\lambda)$ does not even need to be a continuous distribution and $P(\tau)$ could be asymptotically power law for large $\tau$.
In this more general context, we still recover a universal infection kernel $\theta_m(\rho) \propto \rho^\nu$ in most cases, but $\nu$ is not always directly equal to $\alpha$ \cite{SM}.
Linear infection kernels $(\nu = 1)$ are recovered only in some specific cases, for instance when $\alpha = 1$ or when we use a Poisson distribution for $\kappa$ and \mbox{$K = 1$}.
From now on, we make abstraction of the underlying distributions $\pi(\kappa ; \lambda)$ and $P(\tau)$, and focus on the resulting effective model parametrized by $\nu$.

\paragraph{Epidemic spreading with nonlinear infection kernel.}
We now illustrate the consequences of nonlinear infection kernels for contagion on hypergraphs.
To simplify the mathematical analysis, we consider a Susceptible-Infectious-Susceptible model.
At each time step, an infected node becomes susceptible with probability $\mu$, while a susceptible node gets infected through a hyperedge of size $m$ with probability $\theta_m$.
If a node is part of multiple hyperedges, we assume that the probability of infection through each hyperedge are independent.
This is reasonable if \mbox{$k_\mathrm{max} \mathcal{T} \leq \tau_\mathrm{max}$} for a maximal hyperdegree $k_\mathrm{max}$.
For instance, if $\mathcal{T}$ is of the order of a few hours, $\tau_\mathrm{max}$ a week, and an individual participate to an environment once a day, the night allows the immune system to clear any dose $\kappa < K$ accumulated the day before.

To obtain some analytical insights, we introduce a degree-based mean-field theory~\cite{pastor-satorras2015epidemic}, an approximation equivalent to consider an infinite size \textit{annealed} hypergraph, where at each time step, the connections between nodes and hyperedges are shuffled.
We validate our approach with Monte Carlo simulations on \textit{quenched} hypergraphs in the supplemental material~\cite{SM}.

With the mean-field approximation, the marginal probability for a node to be infected at time $t$ only depends on its hyperdegree, which we note $\rho_k(t)$.
The global prevalence is then simply $I(t) = \sum_k \rho_k(t) \tilde{P}(k)$ and the evolution of the system is described by~\cite{pastor-satorras2015epidemic}
\begin{align}
    \rho_k(t+1) = (1-\mu)\rho_k(t)+[1-\rho_k(t)]\Theta_k(\bar{\rho})\;, \label{eq:rhok_time_evo}
\end{align}
where $\Theta_k(\bar{\rho}) = 1 - [1 - \bar{\theta}(\bar{\rho})]^k$ is the probability for a susceptible node of hyperdegree $k$ to get infected. $\bar{\rho}(t)$ is the probability that a node belonging in any hyperedge is infected and $\bar{\theta}(\bar{\rho})$ is the probability for a susceptible node to get infected in any hyperedge,
\begin{align}
    \bar{\rho}(t) = \sum_{k} \rho_k(t) \frac{k \tilde{P}(k)}{\avg{k}} \ \ \ \textrm{and} \ \ \ \bar{\theta}(\bar{\rho}) = \sum_m \bar{\theta}_m (\bar{\rho}) \frac{m \hat{P}(m)}{\langle m \rangle}  \;, \label{eq:rhobar_def}
\end{align}
where $\bar{\theta}_m (\bar{\rho})$ is the probability for a node to get infected in a hyperedge of size $m$. Because of the annealed structure, $\bar{\theta}_m (\bar{\rho})$ is just the average of $\theta_m(\rho)$ with $\rho = i/(m-1)$ over a binomial distribution,
\begin{align}
    \label{eq:thetam_bar_def}
    \bar{\theta}_m(\bar{\rho}) =  \sum_{i =0}^{m-1} \binom{m-1}{i} \bar{\rho}^i (1-\bar{\rho})^{m-1-i} \theta_m \left( \frac{i}{m-1} \right) \;,
\end{align}
with $\theta_m (\rho)$ defined at Eq.~\eqref{eq:thetam_def}.

Figure~\ref{fig:kernel}(b) shows a first consequence of the nonlinear kernel: the apparition of super-exponential growth for the global prevalence $I(t)$ when $\nu > 1$. Note that the growth is approximately exponential until a sufficiently high prevalence.
For $\nu \leq 1$, we instead have a standard exponential growth until saturation is reached.

In the steady state of the epidemic dynamics, we obtain a self-consistent solution
\begin{align}
\rho_k^*=\frac{\Theta_k^*}{\mu+\Theta_k^*} \ \ \  \textrm{and} \ \ \ \bar{\rho}^*=\sum_{k}\tilde{P}(k)\frac{k}{\avg{k}}\frac{\Theta_k^*}{\mu+\Theta_k^*} \equiv G(\bar{\rho}^*) \;,
\label{eq:self-consistent}
\end{align}
since $\Theta_k$ is a function of $\bar{\rho}$.

For contagions with a nonlinear infection kernel, the phase transition associated with the order parameter $I^*$ can be continuous or discontinuous with a bistable regime. Consequently, we define the \textit{invasion} threshold $\beta_\mathrm{c}$ such that for all $\beta > \beta_\mathrm{c}$, the absorbing state $I^* = 0$ is unstable [dashed line in Fig.~\ref{fig:kernel}(c)]. We also define the \textit{persistence} threshold $\beta_\mathrm{p}$ such that for all $\beta < \beta_\mathrm{p}$, the absorbing state $I^* = 0$ is globally attractive [dotted line in Fig.~\ref{fig:kernel}(c)]. For continuous phase transitions, $\beta_\mathrm{c}$ and $\beta_\mathrm{p}$ coincide, and is called the epidemic threshold; for a discontinuous phase transition, $\beta_\mathrm{p} < \beta_\mathrm{c}$, and for all $\beta \in (\beta_\mathrm{p},\beta_\mathrm{c})$, there exists typically three solutions, $I_1^* = 0$ and $I_2^*,I_3^* > 0$, with $I_1^*$ and $I_3^*$ locally stable.

The invasion threshold $\beta_\mathrm{c}$ can be found by imposing $G^{\prime}(0)=1$, the persistence threshold $\beta_\mathrm{p}$ is obtained by imposing both $\bar{\rho}^{*}=G(\bar{\rho}^{*})$ and $G^{\prime}(\bar{\rho}^{*})=1$ for $\bar{\rho}^* > 0$, and any tricritical point can be found by imposing $G^{\prime}(0)=1$ and $G^{\prime\prime}(0)=0$.

In the supplemental material \cite{SM}, we obtain an exact self-consistent expression for the invasion threshold, and using again an asymptotic approximation, we find
\begin{align}
    \label{eq:asymptotic_invasion_threshold}
    \beta_\mathrm{c} \propto \left ( \frac{\mu \left \langle m \right \rangle \left \langle k \right \rangle}{\left \langle m(m-1)^{1-\nu} \left [f(m) \right]^\nu \right \rangle \left \langle k^2 \right \rangle} \right)^{1/\nu} \;.
\end{align}
It depends in a intricate manner on both the moments of $\hat{P}(m)$ and $\tilde{P}(k)$ and the nonlinearity of the infection kernel.
As illustrated in Fig.~\ref{fig:kernel}(c), the invasion threshold can become very small for $\nu < 1$ (vanishing for $\nu \to 0$), even for homogeneous $\tilde{P}(k)$ and $\hat{P}(m)$.
Note that to obtain a sub-linear kernel $\nu < 1$, it typically requires a distribution $P(\tau) \propto \tau^{-\alpha-1}$ with $\alpha < 1$, which implies that the mean participation time $\langle \tau \rangle$ diverges. In the supplemental material \cite{SM}, we show that if instead we fix $\langle \tau \rangle$ while varying $\alpha$, there exists an optimal temporal heterogeneity $\alpha^*$ that minimizes the invasion threshold $\beta_\mathrm{c}$, and maximizes early spread.

The minimal kernel exponent $\nu_\mathrm{c}$ leading to a discontinuous phase transition is given by a tricritical point \cite{SM}.
Although exact solutions require numerical evaluation, we get three insights from an asymptotic expansion.
\begin{enumerate*}[label=(\textit{\roman*})]
    \item $\nu > 1$ is necessary in order to have a discontinuous phase transition, but it is not sufficient: $\nu_\mathrm{c}$ depends on the first three moments of $\tilde{P}(k)$, and in a more complicated manner on the distribution $\hat{P}(m)$.
    \item It is necessary to have environments of size $m > 2$ to have a discontinuous phase transition.
    \item A more heterogeneous $\tilde{P}(k)$ leads to a larger $\nu_\mathrm{c}$.
\end{enumerate*}
Similar observations were made in Ref.~\cite{landry2020effect} for $m \leq 3$.

\paragraph{Conclusion.}
Our framework captures many properties usually overlooked for the sake of simplicity in epidemic models: the higher-order structure of contacts, the temporal heterogeneity of human activity and thresholding effects over the exposure due to the host immune system. In the supplemental material \cite{SM}, we also demonstrate that our results are robust to variations in individual infectiousness or local transmission in different environments.
In particular, we recover a universal nonlinear infection kernel that provides a connection between complex contagions based on nonlinear infection kernels \cite{Liu1987} and threshold models \cite{granovetter1978threshold,watts2002simple,dodds2004universal}.

Our results challenge a key assumption of most epidemic models and ask: Why assume a linear relationship between the number of infectious contacts and the risk of infection?
This question is critical since three of the basic insights gathered from epidemic models break down under nonlinear infection kernels: they can lead to a discontinuous relationship between disease transmission and epidemic size, to a bistable regime where macroscopic outbreak and disease-free state coexist, and to a super-exponential growth.
While the first two are difficult to assess for real contagions, super-exponential spread has been observed for influenza-like illness \cite{scarpino2016effect}.

Even though we considered the SIS model to simplify the analysis, the universal infection kernel $\theta_m(\rho) \propto \rho^\nu$ could be directly integrated in more realistic models---such as SEIR or SIRS---where the same phenomenology typically carries over.

The phenomenology being drastically different from standard epidemiological models begs the following question: Why do linear models work?
Even for a nonlinear kernel $\theta_m(\rho)$, the probability of infection $\bar{\theta}_m(\bar{\rho})$ (averaged over hyperedge configurations) is \textit{linear} in $\bar{\rho}$ if $\bar{\rho} \ll 1$ [see Eq.~\eqref{eq:thetam_bar_def}].
Therefore, linear models are a good approximation when the prevalence is sufficiently low, but breaks down at higher prevalence, as clearly illustrated in Fig.~\ref{fig:kernel}(b) when $\nu = 1.5$.

The mathematical framework we use to solve the SIS model hinges on a mean-field or annealed approximation, as in other studies \cite{Iacopini2019,jhun2019simplicial,ferrazdearruda2020social,landry2020effect}, thereby suppressing \textit{dynamical correlations} within hyperedges.
As we show in the supplemental material \cite{SM}, dynamical correlations can be captured using approximate master equations \cite{hebert2010propagation,marceau2010adaptive,Gleeson2011,Lindquist2011,osullivan2015mathematical,st-onge2021social,st-onge2021master}, which are more complicated but provide similar results with better agreement to simulations.
Future works could investigate more thoroughly the interplay between dynamical correlations, nonlinear kernels, and spatiotemporal heterogeneity.

Altogether, our conclusions stress the need to embrace heterogeneity in disease modeling; in the infection dynamics itself, in patterns of temporal activity, and in the higher-order structure of contact networks. Epidemics should be seen as the result of a collective process, where higher-order structure and temporal patterns can drive complex dynamics.

\begin{acknowledgments}
The authors thank Dr.~Sean~Diehl for help with advanced immunological concepts. This work was supported by the Fonds de recherche du Qu\'{e}bec – Nature et technologies (G.S.), the Natural Sciences and Engineering Research Council of Canada (G.S., A.A.), the Sentinelle Nord program from the Canada First Research Excellence Fund (G.S., A.A.), the Chinese Scholarship Council (H.S.), the National Institutes of Health 1P20 GM125498-01 Centers of Biomedical Research Excellence Award (L.H.-D.), the Royal Society (IEC\textbackslash NSFC\textbackslash191147; G.B.).
\end{acknowledgments}

\clearpage


\section*{Supplementary material}

# Universal nonlinear infection kernel from heterogeneous exposure on higher-order networks
— Supplemental Material —

Guillaume St-Onge,[1,2] Hanlin Sun,[3] Antoine Allard,[1,2] Laurent Hébert-Dufresne,[1,4,5] and Ginestra Bianconi[3,6]

[1]*Département de physique, de génie physique et d'optique,*
*Université Laval, Québec (Québec), Canada G1V 0A6*
[2]*Centre interdisciplinaire en modélisation mathématique,*
*Université Laval, Québec (Québec), Canada G1V 0A6*
[3]*School of Mathematical Sciences, Queen Mary University of London, London, E1 4NS, United Kingdom*
[4]*Vermont Complex Systems Center, University of Vermont, Burlington, VT 05405*
[5]*Department of Computer Science, University of Vermont, Burlington, VT 05405*
[6]*The Alan Turing Institute, 96 Euston Rd, London NW1 2DB, United Kingdom*

## CONTENTS

## I. ASYMPTOTIC NOTATION

In the following sections, we make use of the following notations for asymptotic analysis:

- $f \sim g$ stands for $f$ is asymptotically equal to $g$
- $f = O(g)$ means $f$ is bounded above by $g$
- $f = \Omega(g)$ means $f$ is bounded below by $g$
- $f \ll g$ indicates that $f$ is dominated by $g$
- $f \gg g$ indicates that $f$ dominates $g$

The last two would be the equivalent to the small O and the small Omega in the Bachmann–Landau notation. When $f$ is asymptotically proportional to $g$, we use $f \propto g$, and it should be clear from the context if it is only true in some regime.

## II. NONLINEAR INFECTION KERNEL WITH GENERAL DOSE DISTRIBUTION

As in the main text, let us assume that a susceptible individual participates to an environment of size $m$, with a fraction $\rho$ of the $m-1$ individual being infected, for a duration $\tau$. During this period, the susceptible individual receives an infective dose $\kappa$ of distribution $\pi(\kappa; \lambda)$, where $\lambda \equiv \langle \kappa \rangle$ is

$$\lambda = \beta f(m)\tau\rho . \tag{S1}$$

The susceptible individual develops the infection if $\kappa \geq K$, hence with probability

$$\bar{\Pi}(K; \lambda) = \int_K^\infty \pi(\kappa; \lambda) \, d\kappa . \tag{S2}$$

Note that the distribution could be discrete or continuous.

The infection kernel is then obtained by averaging over the participation time,

$$\theta_m(\rho) = \int_{\mathcal{D}_\tau} \bar{\Pi}(K; \lambda) P(\tau) d\tau , \tag{S3}$$

for some domain $\mathcal{D}_\tau$. In this paper we focus on heterogeneous exposure modeled by $P(\tau) = C_\alpha \tau^{-\alpha-1}$ with $\tau \in [1, \tau_{\max}]$.

As mentioned in the main text, $\bar{\Pi}(K; \lambda)$ is ill-defined for $\tau \geq \mathcal{T}$, where $\mathcal{T}$ is the clearing window for the immune system, but fortunately, we will be able to neglect these events. In fact, we always have that $\bar{\Pi}(K; \lambda) \leq 1$, hence the remainder associated to participation times $\tau \geq \mathcal{T}$ is upper-bounded

$$\mathcal{R}(\mathcal{T}) \leq \int_\mathcal{T}^{\tau_{\max}} P(\tau) d\tau = C_\alpha \int_\mathcal{T}^{\tau_{\max}} \tau^{-\alpha-1} d\tau = O(\mathcal{T}^{-\alpha}) , \tag{S4}$$

As it will become clear in this section, terms $O(\mathcal{T}^{-\alpha})$ have a negligible contribution to the infection kernel. For mathematical convenience, we thus redefine our support as $\tau \in [1, \mathcal{T}]$ and the infection kernel as

$$\theta_m(\rho) = C_\alpha \int_1^\mathcal{T} \bar{\Pi}(K; K\tau/\tau_c) \tau^{-\alpha-1} d\tau , \tag{S5}$$

where, as in the main text, $\tau_c \equiv K/\beta f(m)\rho$ is the characteristic time for an individual to become infected. Integrating by parts, we have

$$\theta_m(\rho) = \frac{C_\alpha}{\alpha} \left[ \bar{\Pi}(K; K/\tau_c) - \bar{\Pi}(K; K\mathcal{T}/\tau_c) \mathcal{T}^{-\alpha} + \int_1^\mathcal{T} \partial_\tau \bar{\Pi}(K; K\tau/\tau_c) \tau^{-\alpha} d\tau \right] . \tag{S6}$$

Making the change of variable $x = \tau/\tau_c$, we obtain

$$\theta_m(\rho) = \frac{C_\alpha}{\alpha} \left[ \bar{\Pi}(K; K/\tau_c) - \bar{\Pi}(K; K\mathcal{T}/\tau_c) \mathcal{T}^{-\alpha} + \tau_c^{-\alpha} \int_{1/\tau_c}^{\mathcal{T}/\tau_c} \partial_x \bar{\Pi}(K; Kx) x^{-\alpha} dx \right] . \tag{S7}$$

As we will make clear with a few cases, in the regime where $1 \ll \tau_c \ll \mathcal{T}$, depending on the asymptotic properties of each term in Eq. (S7), many distributions $\pi(\kappa; \lambda)$ map to the universal nonlinear infection kernel

$$\theta_m(\rho) \sim D_\alpha \tau_c^{-\nu} \propto \rho^\nu , \tag{S8}$$

for some proportionality constant $D_\alpha$.

### A. Dose distribution where $\lambda$ is a scale parameter

An important subclass is all distributions $\pi(\kappa; \lambda)$ where $\lambda$ is a *scale parameter*, i.e.,

$$\pi(\kappa; \lambda) = \frac{1}{\lambda} \pi\left(\frac{\kappa}{\lambda}; 1\right) \equiv \frac{1}{\lambda} \pi\left(\frac{\kappa}{\lambda}\right) ,$$

$$\bar{\Pi}(K; \lambda) = \bar{\Pi}\left(\frac{K}{\lambda}; 1\right) \equiv \bar{\Pi}\left(\frac{K}{\lambda}\right) .$$

In this case, Eq. (S7) becomes

$$\theta_m(\rho) = \frac{C_\alpha}{\alpha} \left[ \bar{\Pi}(\tau_c) - \bar{\Pi}(\tau_c/\mathcal{T})\mathcal{T}^{-\alpha} + \tau_c^{-\alpha} \int_{1/\tau_c}^{\mathcal{T}/\tau_c} \partial_x \bar{\Pi}(1/x) x^{-\alpha} dx \right] . \tag{S9}$$

Making the change of variable $y = 1/x$, and recall that $\bar{\Pi}$ is the *complementary* cumulative distribution, we obtain

$$\theta_m(\rho) = \frac{C_\alpha}{\alpha} \left[ \bar{\Pi}(\tau_c) - \bar{\Pi}(\tau_c/\mathcal{T})\mathcal{T}^{-\alpha} + \tau_c^{-\alpha} \int_{\tau_c/\mathcal{T}}^{\tau_c} \pi(y) y^\alpha dy \right] . \tag{S10}$$

In the regime $1 \ll \tau_c \ll \mathcal{T}$, the integral on the right is bounded from below by some constant, or in asymptotic notation it is $\Omega(1)$, hence the last term is $\Omega(\tau_c^{-\alpha})$, whereas the second term is $O(\mathcal{T}^{-\alpha})$ so we can neglect it. This also justifies neglecting participation times $\tau \geq \mathcal{T}$ since $\mathcal{R}(\mathcal{T}) = O(\mathcal{T}^{-\alpha})$. The infection kernel becomes

$$\theta_m(\rho) \sim \frac{C_\alpha}{\alpha} \left[ \bar{\Pi}(\tau_c) + \tau_c^{-\alpha} \int_{\tau_c/\mathcal{T}}^{\tau_c} \pi(y) y^\alpha dy \right] . \tag{S11}$$

It is now clear that for all distributions $\pi(y)$ that decrease faster than $\tau_c^{-\alpha-1}$ in the tail, the integral converges to a constant $D_\alpha$ and $\bar{\Pi}(\tau_c) \ll \tau_c^{-\alpha}$, therefore, we recover the infection kernel

$$\theta_m(\rho) \sim D_\alpha \tau_c^{-\alpha} , \tag{S12}$$
$$\propto \rho^\alpha . \tag{S13}$$

If instead $\pi(y)$ decreases slower than $y^{-\alpha-1}$, let's say as $y^{-\xi-1}$ for large $y$, where $\xi < \alpha$, then both terms are asymptotically proportional to $\tau_c^{-\xi}$,

$$\theta_m(\rho) \sim E_\alpha \tau_c^{-\xi} , \tag{S14}$$
$$\propto \rho^\xi , \tag{S15}$$

where $E_\alpha$ is another constant prefactor. If $\xi = \alpha$, we would have

$$\theta_m(\rho) \propto \rho^\alpha \ln(1/\rho) . \tag{S16}$$

So except for this limit case, we always have

$$\theta_m(\rho) \propto \rho^\nu , \tag{S17}$$

with $\nu = \min(\alpha, \xi)$. This is illustrated in Fig. S1 with the Weibull and the Fréchet (inverse Weibull) distributions.

## B. Other cases

When $\lambda$ is a scale parameter, the properties of the dose distribution are constrained, for instance the fact that $\langle \kappa \rangle = \lambda$, this implies that $\mathrm{Var}(\kappa) = \langle \kappa^2 \rangle - \langle \kappa \rangle^2 \propto \lambda^2$. Here we investigate other scenarios.

Perhaps the simplest case is to consider $\pi(\kappa; \lambda) = \delta(\kappa - \lambda)$, i.e., the dose is no longer a random variable, it is strictly proportional to the time spent in the environment and to the number of infected people. In this case, $\mathrm{Var}(\kappa) = 0$. We directly have

$$\bar{\Pi}(K; K\tau/\tau_c) = H(K\tau/\tau_c - K) = H(\tau - \tau_c) , \tag{S18}$$

where $H(\cdots)$ is the Heaviside function, and therefore

$$\theta_m(\rho) = C_\alpha \int_1^{\mathcal{T}} H(\tau - \tau_c) \tau^{-\alpha-1} d\tau , \tag{S19}$$

$$= C_\alpha \int_{\tau_c}^{\mathcal{T}} \tau^{-\alpha-1} d\tau \quad \text{if } 1 \leq \tau_c , \tag{S20}$$

$$\sim \frac{C_\alpha}{\alpha} \tau_c^{-\alpha} \quad \text{if } \tau_c \ll \mathcal{T} , \tag{S21}$$

$$\propto \rho^\alpha . \tag{S22}$$

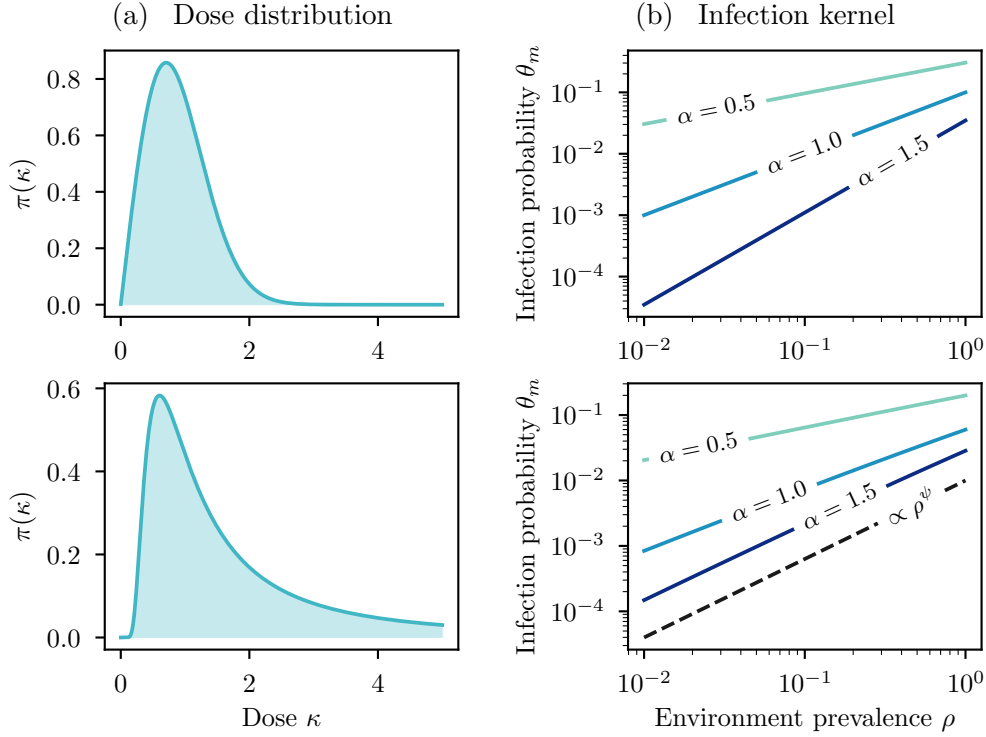

FIG. S1. Infection kernel resulting from different dose distributions, when considering $\lambda$ to be scale parameter, i.e., $\sigma \propto \lambda$. (a) Dose distributions, respectively the Weibull $\pi(\kappa) \propto x^{\psi-1} e^{-(\kappa/\sigma)^{\psi}}$ and the Fréchet $\pi(\kappa) \propto x^{-\psi-1} e^{-(\kappa/\sigma)^{-\psi}}$ distribution from top to bottom. We use $\psi = 2$ for the Weibull and $\psi = 1.2$ for the Fréchet, and fix $\sigma = 1$ (on the left) for visual purpose only. (b) We fix $f(m) = 1$, $\beta = 0.1$, $\mathcal{T} \to \infty$, and ensure $\langle \kappa \rangle = \lambda$ by using $\sigma \propto \lambda$. For the Fréchet distribution, $\nu = \min(\alpha, \psi)$, hence for $\alpha = 1.5$, we see that the kernel exponent is $\psi = 1.2$.

We thus recover the same result as before under the less restrictive condition $1 \leq \tau_c \ll \mathcal{T}$. This suggests that if $\pi(\kappa; \lambda)$ is sufficiently concentrated around its mean, our result should hold.

Another case study we consider is when $\mathrm{Var}(\kappa) \propto \lambda^{2-z}$, where $z \in [0, 1]$. We achieve this using the gamma distribution

$$\pi(\kappa; \sigma, \psi) = \frac{\kappa^{\psi-1} e^{-\kappa/\sigma}}{\Gamma(\psi) \sigma^{\psi}} \ . \tag{S23}$$

Since the mean of the distribution is $\langle \kappa \rangle = \psi \sigma$ and the variance is $\mathrm{Var}(\kappa) = \psi \sigma^2$, we can set $\sigma = \lambda^{1-z}$ and $\psi = \lambda^z$, in which case we respect $\langle \kappa \rangle = \lambda$ and have $\mathrm{Var}(\kappa) = \lambda^{2-z}$. We are thus able to interpolate between two cases: $\lambda$ being a scale parameter ($z = 0$), for which we know our result hold, and where the mean and the variance are the same ($z = 1$). We do not provide an analytical proof, but we show in Fig. S2 that the kernel $\theta_m \propto \rho^{\alpha}$ appears to hold for all $z < 1$, whereas for $z = 1$, we have a kernel $\theta_m \propto \rho^{\nu}$ where $\nu = \min(\alpha, 1)$. These again belong to the general class of universal power-law infection kernels.

A last case we consider is when $\pi(\kappa; \lambda)$ is discrete, i.e., $\kappa \in \mathbb{N}_0$. Let us consider more specifically the Poisson distribution

$$\pi(\kappa; \lambda) = \frac{\lambda^{\kappa} e^{-\lambda}}{\kappa!} \ . \tag{S24}$$

This distribution can be justified from a mechanistic perspective: let us assume that individuals from a hyperedge interact at a constant rate $\beta f(m)$. During the period of duration $\tau$, a susceptible individual has on average $\lambda \equiv \beta f(m) \rho \tau$ interactions with infected individuals (doses), and the distribution for $\kappa$ is the Poisson distribution above.

Without loss of generality, we assume that $K \in \mathbb{N}$ is a small positive integer. It will be more convenient to work with $\tilde{\tau}_c \equiv 1/\lambda f(m)\beta = \tau_c/K$, the characteristic time to receive a unit dose. We rewrite Eq. (S7) as

$$\theta_m(\rho) = \frac{C_{\alpha}}{\alpha} \left[ \bar{\Pi}(K; 1/\tilde{\tau}_c) - \bar{\Pi}(K; \mathcal{T}/\tilde{\tau}_c) \mathcal{T}^{-\alpha} + \tilde{\tau}_c^{-\alpha} \int_{1/\tilde{\tau}_c}^{\mathcal{T}/\tilde{\tau}_c} \partial_x \bar{\Pi}(K; x) x^{-\alpha} dx \right] \ . \tag{S25}$$

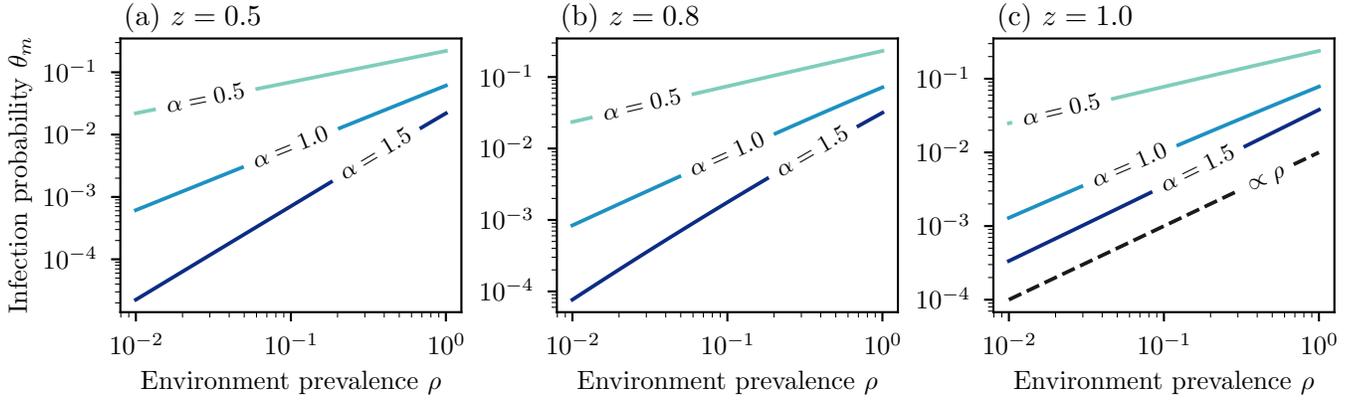

FIG. S2. Infection kernel when $\lambda$ is no longer a scale parameter. We use the gamma distribution at Eq. (S23) with $\sigma = \lambda^{1-z}$ and $\psi = \lambda^z$, with different values of $z$. We use $f(m) = 1, \beta = 0.1$ and $\mathcal{T} \to \infty$.

We have that the complementary cumulative distribution is

$$\bar{\Pi}(K; x) = 1 - \sum_{\kappa=0}^{K-1} \frac{x^\kappa e^{-x}}{\kappa!} , \tag{S26}$$

and

$$\partial_x \bar{\Pi}(K; x) = -\sum_{\kappa=0}^{K-1} \frac{1}{\kappa!} \frac{d}{dx} x^\kappa e^{-x} = e^{-x} + \sum_{\kappa=1}^{K-1} \frac{1}{\kappa!}\left(x^\kappa e^{-x} - \kappa x^{\kappa-1} e^{-x}\right) \tag{S27}$$

For the sake of simplicity, let us consider $K = 2$, which implies

$$\bar{\Pi}(K = 2; 1/\tilde{\tau}_c) = 1 - (1 + \tilde{\tau}_c^{-1})e^{-1/\tilde{\tau}_c} , \tag{S28}$$

$$\sim \frac{\tilde{\tau}_c^{-2}}{2} \quad \text{for } \tau_c \gg 1 , \tag{S29}$$

$$\bar{\Pi}(K = 2; \mathcal{T}/\tilde{\tau}_c) \sim 1 \quad \text{for } \mathcal{T} \gg \tau_c , \tag{S30}$$

and

$$\partial_x \bar{\Pi}(K = 2; x) = xe^{-x} . \tag{S31}$$

Inserting these in Eq. (S25), we obtain

$$\theta_m(\rho) \sim \frac{C_\alpha}{\alpha}\left[\frac{\tilde{\tau}_c^{-2}}{2} + \mathcal{T}^{-\alpha} + \tilde{\tau}_c^{-\alpha} \int_{1/\tilde{\tau}_c}^{\mathcal{T}/\tilde{\tau}_c} x^{1-\alpha} e^{-x} dx\right] . \tag{S32}$$

The integral converges to $\Gamma(2 - \alpha)$ if $1 \ll \tilde{\tau}_c \ll \mathcal{T}$ and $\alpha < 2$. If $\alpha = 2$, the integral diverges as $\ln(\tilde{\tau}_c)$ and for $\alpha > 2$, it diverges as $\tilde{\tau}_c^{\alpha-2}$. Consequently, we have

$$\theta_m(\rho) \propto \begin{cases} \rho^\alpha & \text{if } \alpha < 2 , \\ \rho^2 \ln(1/\rho) & \text{if } \alpha = 2 , \\ \rho^2 & \text{if } \alpha > 2 . \end{cases} \tag{S33}$$

Ignoring the limit case $\alpha = 2$, we have $\theta_m(\rho) \propto \rho^\nu$ where $\nu = \min(\alpha, 2)$. In the general case $K \neq 2$, similarly we can show that $\nu = \min(\alpha, K)$.

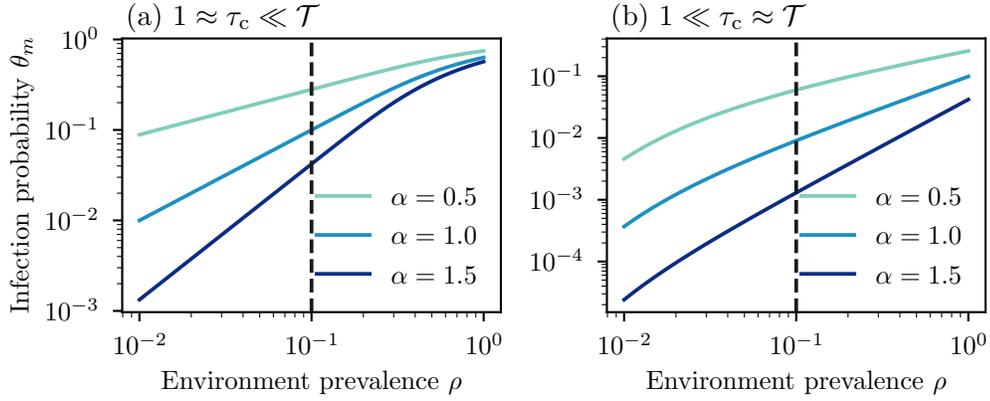

FIG. S3. Infection kernel when the condition $1 \ll \tau_c \ll \mathcal{T}$ is no longer respected. We use an exponential dose distribution $\pi(\kappa; \lambda) = e^{\kappa/\lambda}$ and $f(m) = 1$. (a) $\beta = 1$ and $\mathcal{T} \to \infty$. The vertical dashed line indicates when $\tau_c = 10$. (b) $\beta = 0.1$ and $\mathcal{T} = 10^3$. The vertical dashed line indicates when $\tau_c = 100 = 0.1\mathcal{T}$. Note that $\tau_c$ increases when $\rho$ decreases.

### C. Beyond a pure power-law distribution for participation time

Our previous results assume that $P(\tau)$ is a pure power-law distribution. However, we can relax this assumption by assuming that $P(\tau) \propto \tau^{-\alpha}$ for $\tau > \tau^*$. In this case, it is straightforward to generalize our results by splitting the domain $[1, \mathcal{T}]$ in two parts, $[1, \tau^*]$ and $[\tau^*, \mathcal{T}]$. Equation (S5) becomes

$$\theta_m(\rho) = \int_1^{\tau^*} \bar{\Pi}(K; K\tau/\tau_c) P(\tau) d\tau + C_\alpha \int_{\tau^*}^{\mathcal{T}} \bar{\Pi}(K; K\tau/\tau_c) \tau^{-\alpha-1} d\tau , \qquad (S34)$$

where $C_\alpha$ is again a normalization factor. Note that the first term can be upper-bounded,

$$\int_1^{\tau^*} \bar{\Pi}(K; K\tau/\tau_c) P(\tau) d\tau \leq (\tau^* - 1) M \bar{\Pi}(K; K\tau^*/\tau_c) , \qquad (S35)$$

where $M = \max \{P(\tau) | \tau \in [1, \tau^*]\}$ is some constant. Consequently, Eq. (S34) becomes

$$\theta_m(\rho) = O\left[\bar{\Pi}(K; K\tau^*/\tau_c)\right] + C_\alpha \int_{\tau^*}^{\mathcal{T}} \bar{\Pi}(K; K\tau/\tau_c) \tau^{-\alpha-1} d\tau , \qquad (S36)$$

Integrating by parts the second term, we obtain

$$\theta_m(\rho) = \frac{C_\alpha}{\alpha} \left[ O\left[\bar{\Pi}(K; K\tau^*/\tau_c)\right] - \bar{\Pi}(K; K\mathcal{T}/\tau_c) \mathcal{T}^{-\alpha} + \tau_c^{-\alpha} \int_{\tau^*/\tau_c}^{\mathcal{T}/\tau_c} \partial_x \bar{\Pi}(K; Kx) x^{-\alpha} dx \right] . \qquad (S37)$$

Therefore, noting the similarity with Eq. (S7), all our previous results hold in the more restrictive regime $\tau^* \ll \tau_c \ll \mathcal{T}$.

Instead of having a piecewise distribution $P(\tau)$, many real distributions are asymptotically power law, i.e., $P(\tau) \propto \tau^{-\alpha-1}$ for sufficiently large $\tau$. One just have to pick a $\tau^*$ large enough such that $P(\tau)$ is a sufficiently good approximation of a power law for $\tau > \tau^*$.

### D. Breaking the conditions of the asymptotic approximation

For the asymptotic developments above, we assume that $1 \ll \tau_c \ll \mathcal{T}$. In Fig. S3, we illustrate what happens when this is not respected. For the sake of simplicity, let us say that the conditions are broken if $\tau_c \leq 10$ or $\mathcal{T}/\tau_c \leq 10$.

Let us recall that $\tau_c = 1/\beta f(m)\rho$. In Fig. S3(a), we pick $\beta$ and $f(m)$ such that $\tau_c \leq 10$ for $\rho \geq 0.1$ (on the right of the vertical dashed line), in which case the condition $\tau_c \gg 1$ is not respected. We see that the scaling $\rho^\alpha$ is lost in this regime, but is still valid for $\rho \lesssim 0.1$. In Fig. S3(b), we pick $\mathcal{T}$, $\beta$ and $f(m)$ such that $\mathcal{T}/\tau_c \leq 10$ for $\rho \leq 0.1$ (on the left of the vertical dashed line), in which case the condition $\mathcal{T} \gg \tau_c$ is not respected. We see that the scaling $\rho^\alpha$ is lost in this regime, but is still valid for $\rho \gtrsim 0.1$.

In summary, the condition $\mathcal{T} \gg \tau_c$, for which we can choose a specific tolerance such as $\mathcal{T}/\tau_c \geq 10$, gives a lower bound $\rho_1$ for the validity of $\theta_m(\rho) \propto \rho^\alpha$, while the condition $\tau_c \gg 1$, or as we chose $\tau_c \geq 10$, gives an upper bound $\rho_2$. We can therefore assess that the infection kernel is valid on some domain $[\rho_1, \rho_2]$.

### E. Robustness to other sources of heterogeneity

The results put forward in this paper highlight the role of the burstiness of interactions for the emergence of nonlinear infection kernels. The heterogeneity of degree $k_r$ and environment sizes $m_\gamma$ is also incorporated in the modeling approach we use in the main text. In this section, we show that adding other layers of heterogeneity do not alter the form of the universal kernel $\theta_m(\rho) \propto \rho^\nu$. We demonstrate it in the case where $\lambda$ is a scaling parameter for the sake of simplicity.

#### 1. Heterogeneity of environments

The attack rate in an environment is certainly affected by the conditions of the environment itself: for instance, non-ventilated places might be more prone to infections than better ventilated ones. Let us assume that $\beta \in [0, \infty)$ is an i.i.d random variable for each environment, drawn from an arbitrary distribution $Q(\beta)$.

If $Q(\beta)$ is a power law, it plays a similar role as $P(\tau)$, since $\beta$ and $\tau$ appear side by side in the definition of the mean dose $\langle \kappa \rangle = \lambda$. With $\tau$ and $\beta$ both distributed as power laws, the distribution $R(x)$ for $x = \tau\beta$ is a mixture of power-law distributions [1]. The tail of $R(x)$ is then dominated by the slowest decaying distribution, either $Q(\beta)$ or $P(\tau)$. In all cases, we could extend the analysis above using $R(x)$ instead $P(\tau)$, and we would recover that $\theta_m(\rho) \propto \rho^\nu$ with $\nu$ being the exponent of the distribution that decreases the slowest (see Fig. S4 lower row). We therefore assume that $Q(\beta)$ decreases faster than a power law, i.e., $Q(\beta) \ll \beta^{-n}\ \forall n \in \mathbb{N}$ in the limit $\beta \to \infty$.

Recall that $\tau_c^{-1} \equiv f(m)\beta\rho$, hence $\tau_c = \tau_c(\beta)$. We can rewrite Eq. (S10) as

$$\theta_m(\rho) = \int_0^\infty \frac{C_\alpha}{\alpha}\left[\bar{\Pi}(\tau_c) - \bar{\Pi}(\tau_c/\mathcal{T})\mathcal{T}^{-\alpha} + \tau_c^{-\alpha}\int_{\tau_c/\mathcal{T}}^{\tau_c}\pi(y)y^\alpha \mathrm{d}y\right]Q(\beta)\mathrm{d}\beta \ . \tag{S38}$$

Let now us suppose that we can divide the support for $\beta$ in three subintervals, $\mathcal{B}_1 \equiv [0, \beta_1]$, $\mathcal{B}_2 \equiv [\beta_1, \beta_2]$ and $\mathcal{B}_3 \equiv [\beta_2, \infty)$, with $\int_{\beta \in \mathcal{B}_2} Q(\beta)\mathrm{d}\beta$ sufficiently large. In the middle subinterval, we assume that the usual condition for asymptotic expansion holds, $1 \ll \tau_c(\beta) \ll \mathcal{T}$ for all $\beta \in \mathcal{B}_2$. In the other subintervals, this condition is not respected for all $\beta$, i.e., there are values for $\beta \in \mathcal{B}_1$ such that $\tau_c(\beta) \approx \mathcal{T}$ or even $\tau_c(\beta) \gg \mathcal{T}$ and values for $\beta \in \mathcal{B}_2$ such that $\tau_c(\beta) \approx 1$ or even $\tau_c(\beta) \ll 1$. We investigate the asymptotic behavior of $\theta_m^{\mathcal{B}_j}$ for each subinterval $\mathcal{B}_j$, where $\theta_m = \sum_{j=1}^3 \theta_m^{\mathcal{B}_j}$.

In the subinterval $\mathcal{B}_2$, we can replace the inside of the integral on $\beta$ with the usual asymptotic expansion, leading to

$$\theta_m^{\mathcal{B}_2}(\rho) \sim D_\alpha \int_{\beta \in \mathcal{B}_2} \tau_c^{-\alpha}(\beta)Q(\beta)\mathrm{d}\beta \ , \tag{S39}$$

$$= D_\alpha \rho^\alpha f(m)^\alpha K^{-\alpha} \int_{\beta \in \mathcal{B}_2} \beta^\alpha Q(\beta)\mathrm{d}\beta \ , \tag{S40}$$

$$\propto \rho^\alpha \ . \tag{S41}$$

In the subinterval $\mathcal{B}_1$, note that smaller values of $\tau_c$ increases the value of the integrand on $\beta$—by definition, $\lambda \propto \tau_c^{-1}$, hence the mean dose received is larger for smaller $\tau_c$. Therefore, we can bound from above $\theta_m^{\mathcal{B}_1}$ by using for all $\beta$ the smallest value possible on the subinterval, $\tau_c(\beta) \mapsto \tau_c(\beta_1) \ll \mathcal{T}$. In this case, the asymptotic expansion we used for $\mathcal{B}_2$ is again possible, which implies

$$\theta_m^{\mathcal{B}_1}(\rho) = O\left[\tau_c^{-\alpha}(\beta_1)\right] \ . \tag{S42}$$

However, Eq. (S39) implies that $\theta_m^{\mathcal{B}_2}(\rho) = \Omega\left[\tau_c^{-\alpha}(\beta_1)\right]$, which suggest that we can have $\theta_m^{\mathcal{B}_1}(\rho) \ll \theta_m^{\mathcal{B}_2}(\rho)$. In fact, this is respected if

$$\beta_1^\alpha \int_{\beta \in \mathcal{B}_1} Q(\beta)\mathrm{d}\beta \ll \int_{\beta \in \mathcal{B}_2} \beta^\alpha Q(\beta)\mathrm{d}\beta \ , \tag{S43}$$

in which case we can neglect the contribution from $\theta_m^{\mathcal{B}_1}(\rho)$

Finally, in the subinterval $\mathcal{B}_3$, we again use the fact that smaller values of $\tau_c$ increases the value of the integrand on $\beta$. In the limit $\beta \to \infty$, $\tau_c \to 0$, and we are left with the upper bound

$$\theta_m^{\mathcal{B}_3}(\rho) \leq \int_{\beta \in \mathcal{B}_3} Q(\beta)\mathrm{d}\beta \ . \tag{S44}$$

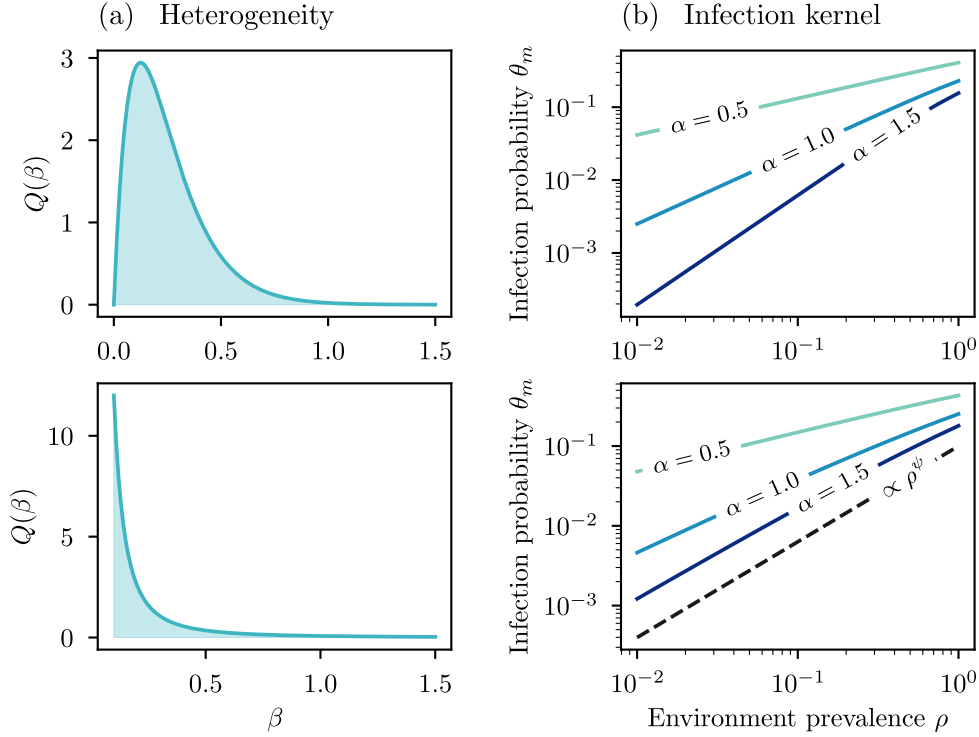

FIG. S4. Nonlinear infection kernels are robust to heterogeneous infectiousness. (a) Distributions for the heterogeneity of the environment. On the upper row, we use a gamma distribution $Q(\beta) \propto \beta^{\psi-1} e^{-\beta/\sigma}$ with $\sigma = 1/8$, $\psi = 2$, and $\beta \in [0, \infty)$. On the lower row, we use a power-law distribution $Q(\beta) \propto \beta^{-\psi}$ with $\psi = 1.2$ and $\beta \in [0.5, \infty)$. (b) Infection kernel obtained using the distribution $Q(\beta)$ on the left. We use $\pi(\kappa; \lambda) \propto e^{-\kappa/\lambda}$, $P(\tau) \propto \tau^{-\alpha-1}$, $\mathcal{T} \to \infty$ and $f(m) = 0.1$. For the gamma distribution (upper row), we have $\theta_m(\rho) \propto \rho^\alpha$, while for the power-law distribution (lower row), we have $\theta_m(\rho) \propto \rho^\nu$ with $\nu = \min(\alpha, \psi)$ and $\psi = 1.2$.

For the universal kernel $\theta_m(\rho) \propto \rho^\alpha$ to hold, we must have

$$\theta_m^{\mathcal{B}_3}(\rho) \ll \theta_m^{\mathcal{B}_2}(\rho) \,. \tag{S45}$$

A sufficient condition is if

$$\int_{\beta \in \mathcal{B}_3} Q(\beta) \mathrm{d}\beta \ll D_\alpha \rho^\alpha f(m)^\alpha K^{-\alpha} \int_{\beta \in \mathcal{B}_2} \beta^\alpha Q(\beta) \mathrm{d}\beta \,. \tag{S46}$$

More intuitively, conditions at Eqs (S43) and (S46) indicate that the probability for $\beta \in \mathcal{B}_1$ and $\beta \in \mathcal{B}_3$ must be sufficiently small, i.e., environments of low or intense spreading are rare compared with environments where the spreading is moderate, satisfying $1 \ll \tau_c(\beta) \ll \mathcal{T}$. Infection kernels obtained with different $Q(\beta)$ are presented in Fig. S4, where we once again recover the form $\theta_m(\rho) \propto \rho^\nu$.

### 2. Heterogeneity of individual infectiousness

Similarly, all infectious individuals are not equivalent. For instance, the viral load within an infectious individual depends on the time since infection. Or more simply, people's behavior varies widely, some follow diligently public health recommendations (hand-washing, social distancing, etc.) and some do not. This motivates assigning i.i.d. random variable $\beta_j$ with distribution $\tilde{Q}(\beta_j)$ to characterize the infectiousness of each infected individual $j \in \{1, \ldots, i\}$, where $i \equiv \rho(m-1)$. A susceptible individual then receives a dose $\kappa_j$ with distribution $\tilde{\pi}(\kappa_j; \lambda_j)$ from each infectious individual $j$, where $\langle \kappa_j \rangle = \lambda_j = \beta_j f(m) \tau \rho$.

However, it is straightforward to verify that adding this additional source of heterogeneity does not affect our results either. If we define $\beta = \sum_{j=1}^{i} \beta_j$, with distribution $Q(\beta)$, and $\lambda = \sum_{j=1}^{i} \lambda_j$, then $\kappa = \sum_{j=1}^{i} \kappa_j$ has a distribution $\pi(\kappa; \lambda)$ corresponding to

$$\pi(\kappa; \lambda) = \frac{1}{Q(\beta)} \int_0^\infty \cdots \int_0^\infty \delta\left(\kappa - \sum_{j=1}^{i} \kappa_j\right) \delta\left(\beta - \sum_{j=1}^{i} \beta_j\right) \prod_{j=1}^{i} \left\{\tilde{\pi}(\kappa_j; \lambda_j) \tilde{Q}(\beta_j) \mathrm{d}\kappa_j \mathrm{d}\beta_j\right\} \,. \tag{S47}$$

Therefore, this is totally equivalent to the case of a heterogeneous environment, with the infection kernel $\theta_m(\rho)$ defined by Eq. (S38).

### 3. Binomial distribution for the participation to hyperedges

In the main text, it is assumed that a node of hyperdegree $k$ interacts in $k$ different environments *at each time step*. More realistically, we can consider that an individual participate to a hyperedge with probability $p$, hence the effective degree of a node for a time window is drawn from a binomial distribution with parameters $k$ and $p$.

This can be incorporated in our approach simply by using

$$P(\tau) = C_\alpha p \tau^{-\alpha-1} + (1-p)\delta(\tau) . \tag{S48}$$

The effect is simply to map $\theta_m(\rho) \mapsto p\theta_m(\rho)$, leaving its scaling with $\rho$ unaltered. Therefore, the whole phenomenology is preserved.

### F. More general temporal patterns

For a specific environment, we suggest in the main text that the participation time $\tau_j$ of each individual $j$ is i.i.d., with distribution $P(\tau)$. However, our result about the infection kernel does not require the independence of the participation time. For instance, the participation time of all individuals belonging to an environment could be perfectly correlated, i.e., $\tau_j = \tau$ for each individual $j$ belonging to the environment. This could be more representative of certain scenarios, for example, of certain workplaces where people have the exact same schedule. More generally, we expect some varying level of correlation for the participation time to different types of environments.

The point is that correlations would be reflected on the form of the dose distribution $\pi(\kappa; \lambda)$ and its mean—a more correlated participation time would be associated with a larger rate of dose accumulation $\beta$ since the individuals spend more time altogether in the environment. Since our result does not depend on the form of $\pi(\kappa; \lambda)$, if the condition $1 \ll \tau_c \ll \mathcal{T}$ is respected, so should be our result about the infection kernel $\theta_m(\rho) \propto \rho^\nu$.

## III. CHARACTERIZATION OF THE PHASE TRANSITION

As discussed in the main text, the phase transition is characterized by $G(\bar{\rho}^*)$. In this section, we derive self-consistent equations for both the invasion threshold $\beta_c$ and any tricritical point $(\beta_c, \nu_c)$, where $\nu_c$ is a limit value for the nonlinear kernel exponent separating a continuous and a discontinuous phase transition.

All time-varying quantities are assumed to have reached the steady state—to simplify the notation, we drop the asterisk.

First, let us rewrite $G(\bar{\rho})$ [Eq. (7) in the main text] and its derivatives as

$$G(\bar{\rho}) = \frac{1}{\langle k \rangle} \left\langle \frac{k\Theta_k}{\mu + \Theta_k} \right\rangle ,$$

$$G'(\bar{\rho}) = \frac{1}{\langle k \rangle} \left\langle \frac{k\Theta'_k(\bar{\rho})}{\mu + \Theta_k} - \frac{k\Theta_k\Theta'_k(\bar{\rho})}{(\mu + \Theta_k)^2} \right\rangle ,$$

$$G''(\bar{\rho}) = \frac{1}{\langle k \rangle} \left\langle \frac{k\Theta''_k(\bar{\rho})}{\mu + \Theta_k} - \frac{2k\left[\Theta'_k(\bar{\rho})\right]^2}{(\mu + \Theta_k)^2} - \frac{k\Theta_k\Theta''_k(\bar{\rho})}{(\mu + \Theta_k)^2} + \frac{2k\Theta_k\left[\Theta'_k(\bar{\rho})\right]^2}{(\mu + \Theta_k)^3} \right\rangle .$$

In the limit $\bar{\rho} \to 0$, we have $\bar{\theta} \to 0$, which implies $\Theta_k \to 0$ for all $k$ and

$$\Theta'_k(0) = k\bar{\theta}'(0) ,$$

$$\Theta''_k(0) = k\bar{\theta}''(0) - k(k-1)\left[\bar{\theta}'(0)\right]^2 .$$

Therefore, we have

$$G'(0) = \frac{\langle k \rangle}{\mu} \bar{\theta}'(0) , \tag{S49a}$$

$$G''(0) = \frac{1}{\mu} \left\{ \langle k^2 \rangle \bar{\theta}''(0) - \left( \langle k^2(k-1) \rangle + \frac{2\langle k^3 \rangle}{\mu} \right)\left[\bar{\theta}'(0)\right]^2 \right\} . \tag{S49b}$$

Second, $\bar{\theta}(\bar{\rho})$ is rewritten as

$$\bar{\theta}(\bar{\rho}) = \frac{1}{\langle m \rangle} \left\langle m\bar{\theta}_m(\bar{\rho}) \right\rangle , \tag{S50}$$

hence it depends on $\bar{\rho}$ only through $\bar{\theta}_m(\bar{\rho})$.

Finally, it is straightforward to obtain the derivatives of Eq. (6) in the main text for $\bar{\rho} = 0$, i.e.,

$$\bar{\theta}'_m(0) = (m-1)\theta_m\left(\frac{1}{m-1}\right) , \tag{S51a}$$

$$\bar{\theta}''_m(0) = (m-1)(m-2)\left[\theta_m\left(\frac{2}{m-1}\right) - 2\theta_m\left(\frac{1}{m-1}\right)\right] . \tag{S51b}$$

Combining Eqs. (S49), (S50) and (S51), we have self-consistent equations to solve for both the invasion threshold and the tricritical points.

### A. Invasion threshold

We can identify the invasion threshold $\beta_c$ for a fixed structure given by $\tilde{P}(k)$ and $\hat{P}(m)$, and for an arbitrary distribution $P(\tau)$. Combining Eqs. (S49a), (S50) and (S51a), the constraint $G'(0) = 1$ becomes

$$\left(\frac{\left\langle m(m-1)\theta_m\left(\frac{1}{m-1}\right)\right\rangle}{\langle m \rangle}\right)\left(\frac{\langle k^2 \rangle}{\langle k \rangle}\right) = \mu . \tag{S52}$$

The left-hand side is intuitively interpreted as the average number of new infections caused by an infected node in a single time step near the absorbing state.

Let us assume that the conditions for the universal infection kernel are respected. Consequently, we can use

$$\theta_m(\rho) \sim D_\alpha \tau_c^\nu , \tag{S53}$$
$$\propto [\beta f(m)\rho]^\nu . \tag{S54}$$

We keep the dependence on $\beta$ and $f(m)$ to see the impact of these parameters, whereas $D_\alpha$ is just some constant that we cannot tune like the others. Consequently, we have $\theta_m(1/m-1) \propto \beta^\nu [f(m)]^\nu (m-1)^{-\nu}$, valid when $\beta f(m)/(m-1) \ll 1$ and $\mathcal{T}\beta f(m)/(m-1) \gg 1$. We obtain

$$\beta_c \propto \left(\frac{\mu \langle m \rangle \langle k \rangle}{\langle m(m-1)^{1-\nu}[f(m)]^\nu \rangle \langle k^2 \rangle}\right)^{1/\nu} . \tag{S55}$$

In the case $f(m) = 1$, we see that having $\nu < 1$ induces a very small value for the invasion threshold.

### B. Tricritical points

For a fixed structure given by $\tilde{P}(k)$ and $\hat{P}(m)$, the tricritical points $(\beta_c, \nu_c)$ are solutions to both $G'(0) = 1$ and $G''(0) = 0$. We substitute $\bar{\theta}'(0) = \mu/\langle k \rangle$ in (S49b) to obtain

$$G''(0) = \frac{1}{\mu}\left\{\langle k^2 \rangle \bar{\theta}''(0) - \left(\langle k^2(k-1) \rangle + \frac{2\langle k^3 \rangle}{\mu}\right)\frac{\mu^2}{\langle k \rangle^2}\right\} . \tag{S56}$$

Since the second term is strictly positive, it is necessary that $\bar{\theta}''(0) > 0$ to have a tricritical point.

Using again the asymptotic expansion for $\theta_m(\rho)$ in combination with Eq. (S50) and (S51b), we get

$$\bar{\theta}''(0) \propto \frac{\beta^\nu}{\langle m \rangle}\left\langle m(m-1)^{1-\nu}(m-2)[f(m)]^\nu (2^\nu - 2)\right\rangle . \tag{S57}$$

The tricritical point is evaluated at $\beta = \beta_c$, hence we can substitute the asymptotic expansion for the invasion threshold [Eq. (S55)] to have an expression that only depends on $\nu$,

$$\bar{\theta}''(0) \propto \mu \frac{\langle k \rangle}{\langle k^2 \rangle} \frac{\left\langle m(m-1)^{1-\nu}(m-2)[f(m)]^\nu (2^\nu - 2)\right\rangle}{\left\langle m(m-1)^{1-\nu}[f(m)]^\nu \right\rangle} . \tag{S58}$$

We get three insights from Eqs. (S56) and (S58).

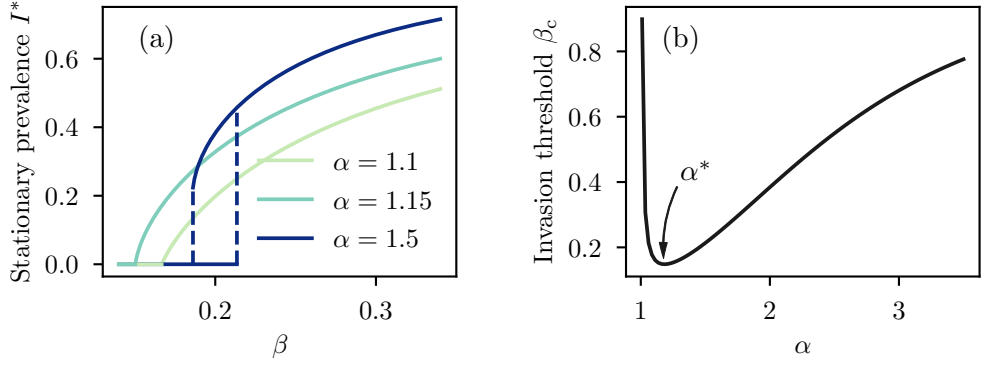

FIG. S5. Impact of a fixed mean participation time $\langle \tau \rangle = 1$. We use a power-law distribution $P(\tau) \propto \tau^{-\alpha-1}$, $\tau \in [(\alpha - 1)/\alpha, \infty)$, $f(m) = 1$, and $\pi(\kappa; \lambda) \propto e^{-\kappa/\lambda}$ with $K = 1$. We use Poisson distributions for both $\tilde{P}(k)$ and $\hat{P}(m)$, with $\langle k \rangle = 5$ and $\langle m \rangle = 10$, and set $\mu = 0.05$. (a) Phase diagram, highlighting the differences compared with a fixed $\tau_{\min}$ as in Fig. 2 in the main text. For a sufficiently large $\beta$, $I^*$ now grows with $\alpha$. The dashed vertical line indicates the position of the persistence and invasion threshold for $\alpha = 1.5$. (b) The invasion threshold $\beta_c$ is minimized for a value $\alpha^* \approx 1.15$. $\beta_c$ goes to infinity in the limit $\alpha \to 1$, and converges to a certain value in the limit $\alpha \to \infty$.

1. It is necessary that $\nu > 1$ to have a tricritical point, but not sufficient: it depends on the first three moments of $\tilde{P}(k)$, and in a more complicated manner on the distribution $\hat{P}(m)$.

2. It is necessary to have $\hat{P}(m) > 0$ for at least one value $m > 2$, i.e., environments of size $m = 2$ cannot lead to a discontinuous phase transition.

3. A more heterogeneous $\tilde{P}(k)$ will typically require a larger $\nu$ to reach a tricritical point. Indeed, if we keep $\langle k \rangle$ fixed, but increase the value for the second and third moments (using a broader distribution for instance), the negative term on the right in Eq. (S56) increases, while the positive term is invariant.

### C. Participation time distribution with a fixed mean

In the main text, we fixed the minimal participation time to $\tau_{\min} = 1$, thereby fixing the scale of events relative to this minimal period, but we let $\langle \tau \rangle$ vary with $\alpha$. Here we show the consequence of fixing $\langle \tau \rangle = 1$, and to let $\tau_{\min}$ vary. For the sake of simplicity, we assume $\mathcal{T} \to \infty$. For a power-law distribution $P(\tau) = C_\alpha \tau^{-\alpha-1}$, this is achieved if we set $\tau_{\min} = (\alpha - 1)/\alpha$. Note that it is only possible for $\alpha > 1$, i.e., when the mean is well-defined.

As shown in Fig. S5, keeping a fixed $\langle \tau \rangle$ reveals another story for the bifurcation diagram. For a sufficiently large $\beta$, we now have that $I^*$ grows with $\alpha$—a homogeneous participation time lead to bigger outbreaks—, which is the opposite of what we found for a fixed $\tau_{\min}$. Moreover, the invasion threshold is now a non-monotonic function of $\alpha$. It is minimized for some value $\alpha^* \approx 1.2$, and the early spread of an epidemic would be maximized for this $\alpha^*$.

## IV. TRACKING DYNAMICAL CORRELATIONS FOR QUENCHED SYSTEMS

In the main text, we supposed an annealed structure—nodes and hyperedges are shuffled at each time step. This is why we can assume that the number of infected individuals in a hyperedge of size $m$ is distributed according to a binomial distribution

$$H_{m,i} = \binom{m}{i} \bar{\rho}^i (1 - \bar{\rho})^{m-i} . \tag{S59}$$

When calculating $\bar{\theta}_m(\bar{\rho})$ at Eq. (6), we restrict ourselves to hyperedges with at least one susceptible node, hence we average $\theta_m[i/(m-1)]$ over the distribution

$$\frac{(m-i)H_{m,i}}{\sum_i (m-i) H_{m,i}} = \frac{(m-i)}{(1-\bar{\rho})m} H_{m,i} = \binom{m-1}{i} \bar{\rho}^i (1-\bar{\rho})^{m-i-1} , \tag{S60}$$

which is again a binomial, but over the $m - 1$ remaining nodes.

In reality, people do engage with the same people in a recurrent manner, which can be represented by a *quenched structure*, i.e., an individual always participate to the same hyperedges. In this case, we can track the evolution of $H_{m,i}(t)$ using approximate master equations (AMEs) of the form

$$H_{m,i}(t+1) = \sum_{l=0}^{m-i} \sum_{j=0}^{i} B_{m+j-i-l,j}(\phi_{m,i+l-j}) B_{i+l-j,l}(\mu) H_{m,i+l-j}(t) \ . \tag{S61}$$

where

$$B_{u,v}(q) = \binom{u}{v} q^v (1-q)^{u-v} \ . \tag{S62}$$

In Eq. (S61), $l$ counts the number of recoveries from $t$ to $t+1$ and $j$ counts the number of infections. $\phi_{m,i} \equiv 1 - (1 - \theta_{m,i})(1 - R)$ is the probability that a susceptible node belonging to a hyperedge $(m,i)$ becomes infected. We have defined

$$\theta_{m,i} \equiv \theta_m \left( \frac{i}{m-1} \right) \ , \tag{S63}$$

which is the probability of getting infected by the hyperedge $(m,i)$, while $R$ is the average probability for a susceptible node to get infected through any *other* hyperedge to which it belongs. It corresponds to

$$R(t) = \sum_k \frac{k[1-\rho_k(t)]\tilde{P}(k)}{\sum_k k[1-\rho_k(t)]\tilde{P}(k)} \left\{ 1 - \left[1 - \bar{\theta}(t)\right]^{k-1} \right\} \ . \tag{S64}$$

The term in curly brace $\{\cdots\}$ is the probability for a susceptible node of hyperdegree $k$ to get infected in any of the other $k-1$ hyperedges to which it belongs. We average all this over a distribution $\propto k\rho_k(t)\tilde{P}(k)$, which is the probability for a random susceptible node in a hyperedge to be of hyperdegree $k$. Finally, the probability $\bar{\theta}$ is now calculated as

$$\bar{\theta}(t) = \frac{\sum_{m,i}(m-i)\theta_{m,i}H_{m,i}(t)\hat{P}(m)}{\sum_{m,i}(m-i)H_{m,i}(t)\hat{P}(m)} \ . \tag{S65}$$

Combining these equations with Eq. (4) in the main text, we have a closed system.

Note that this new set of equations still depends on the infection kernel $\theta_{m,i}$, hence our main result is directly transposed to this case where dynamical correlations are taken into account. In fact, the derivation of the infection kernel is agnostic to the underlying mathematical framework used to model the contagion.

It is also worth pointing out that we do not capture *exactly* all the local dynamical correlations with Eq. (S61): although we track $H_{m,i}(t)$, we assume that each susceptible node gets infected with independent probability $\phi_{m,i}$. However, for $\mu \ll 1$ and $\phi_{m,i} \ll 1$, at most one node either recovers or becomes infected at the same time—in other words, we have asynchronous state transitions. In this limit (effectively taking the continuous time limit), we capture exactly the dynamical correlations within hyperedges. Therefore, Eq. (S61) can be seen as a good approximation for small $\mu$ and $\phi_{m,i}$.

In Fig. S6, we show that the phenomenology of the contagion process using AMEs is very similar, almost indistinguishable from the phenomenology of the annealed system [see Fig. 2(b) and 2(c) in the main text]. Linear and sublinear kernel ($\nu \leq 1$) still yield a standard exponential spread with a continuous phase transition, with a very small invasion threshold $\beta_c$ for small $\nu$. Supralinear kernel ($\nu > 1$) still lead to super-exponential spread with a discontinuous phase transition.

It worth mentioning that even though the results presented here suggest that dynamical correlations within hyperedges have a negligible impact on the contagion process, this does not hold for all scenarios (see Fig. S7). Moreover, in Ref. [2, 3], it was shown that these dynamical correlations are crucial for the onset of mesoscopic localization, which does significantly alter the phenomenology of the system. We expect dynamical correlations to have a larger impact for heterogeneous hyperedge size distribution $\hat{P}(m)$ and more nonlinear contagions (larger $\nu$).

## V. COMPARISON WITH SIMULATIONS

To validate both the annealed equations and the AMEs, we performed simulations of the contagion process on random hypergraphs.

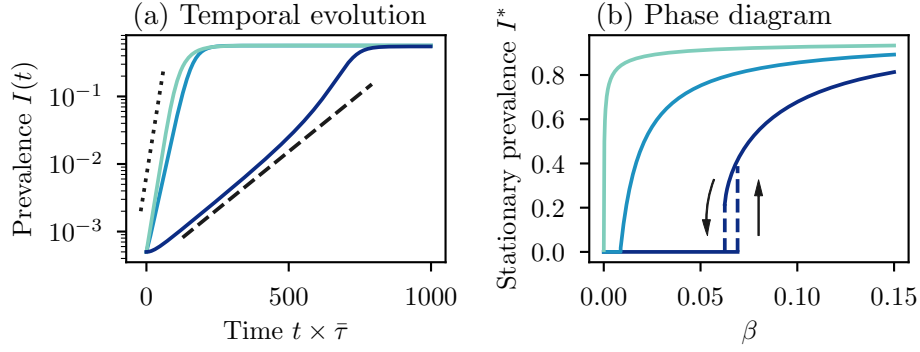

FIG. S6. Temporal evolution and phase diagram using the same conditions as Fig. 2 in the main text, but for a quenched system, making use of approximate master equations [Eqs. (S61)-(S65)].

### A. Generation of random hypergraphs

We generate random hypergraphs with fixed hyperdegree and hyperedge size distributions $\tilde{P}(k)$ and $\hat{P}(m)$. First, we draw a hyperdegree sequence $\boldsymbol{k} = [k_1, k_2, \ldots, k_N]$ from $\tilde{P}(k)$, where the number of nodes $N$ is initially fixed. Second, we draw a hyperedge size sequence $\boldsymbol{m} = [m_1, m_2, \ldots, m_{N'}]$ from $\hat{P}(m)$, where the number of hyperedges $N'$ is not fixed. We want to have $\sum_{j=1}^{N} k_j = \sum_{j=1}^{N'} m_j$, therefore we add hyperedges one at a time until $\sum_{j=1}^{N} k_j \leq \sum_{j=1}^{N'} m_j$. If $\sum_{j=1}^{N} k_j < \sum_{j=1}^{N'} m_j$, we remove hyperedges randomly one at a time until $\sum_{j=1}^{N} k_j \geq \sum_{j=1}^{N'} m_j$. We alternate between these two processes until the equality is reached, which also determines $N'$.

Third, we assign the nodes to the hyperedges, best interpreted as a bipartite graph realization. We create a stub list (list of half edges) for each sequence. For instance, for node $j$, we include $k_j$ copies of the label $j$ in the stub list of nodes. Similarly, we include $m_j$ copies of hyperedge $j$ in the stub list of hyperedges. These two stub lists are of the same size by definition, hence we shuffle both and match the entries of the lists, thereby assigning nodes to hyperedges at random.

However, this process does not avoid assigning a node $j$ multiple times to a same hyperedge if $k_j > 1$. To remove these multiple assignments, we swap the concerned entries of the node stub list with another chosen at random, making sure it does not create another hyperedge with multiple assignments. In the bipartite representation, this move is equivalent to a double edge swap [4, 5]. We then perform a number of additional random double edge swaps ($N$ in our simulations) to ensure the uniformity of the generating process.

### B. Simulation of the contagion process

Let us denote a node $j$ and the set of hyperedges $\gamma$ to which it is assigned as $\partial j$. The parameters $\alpha, T, \beta, f(m)$ and $K$ are fixed. To simulate the contagion process, we start with an initial set of infected nodes $\mathcal{I}(0)$ and a set of susceptible nodes $\mathcal{S}(0)$ at time $t = 0$. At each time step $t \to t + 1$, we perform the following operations.

1. We create a set of *new susceptible* $\tilde{\mathcal{S}}(t + 1)$. For all $j \in \mathcal{I}(t)$, node $j$ is included in set $\tilde{\mathcal{S}}(t + 1)$ with probability $\mu$.

2. We create a set of *new infected* $\tilde{\mathcal{I}}(t + 1)$. For all $j \in \mathcal{S}(t)$, we iterate over hyperedges $\gamma \in \partial j$. For each hyperedge $\gamma$ we:

    (i) draw a participation time $\tau_\gamma$;
    
    (ii) given $\tau_\gamma$, the hyperedge size $m_\gamma$, and the fraction of the other nodes that are infected $\rho_\gamma$, we draw a dose $\kappa_\gamma$ from $\pi(\kappa; \lambda_\gamma)$, where $\lambda_\gamma = \beta f(m) \tau_\gamma \rho_\gamma$;

3. For all susceptible node $j \in \mathcal{S}(t)$, node $j$ is added to the set $\tilde{\mathcal{I}}(t + 1)$ if $\kappa_\gamma > K$ for any of the hyperedge $\gamma \in \partial j$.

4. We update the sets: $\mathcal{I}(t + 1) = \mathcal{I}(t) \bigcup \tilde{\mathcal{I}}(t + 1) \setminus \tilde{\mathcal{S}}(t + 1)$ and $\mathcal{S}(t + 1) = \mathcal{S}(t) \bigcup \tilde{\mathcal{S}}(t + 1) \setminus \tilde{\mathcal{I}}(t + 1)$.

The prevalence at each time step is simply $I(t) = |\mathcal{I}(t)|$.

While the previous algorithm is sufficient to simulate the time evolution of the contagion process, estimating quantities in the stationary state can be problematic for finite size systems because the contagion process can get stuck in the absorbing state where all nodes are susceptible. We therefore use the *quasistationary-state method* [6] to avoid this problem. We keep a history of past states $(\mathcal{I}, \mathcal{S})$—50 states in our case—which we update after each decorrelation period $t_d$. To update the history, we replace

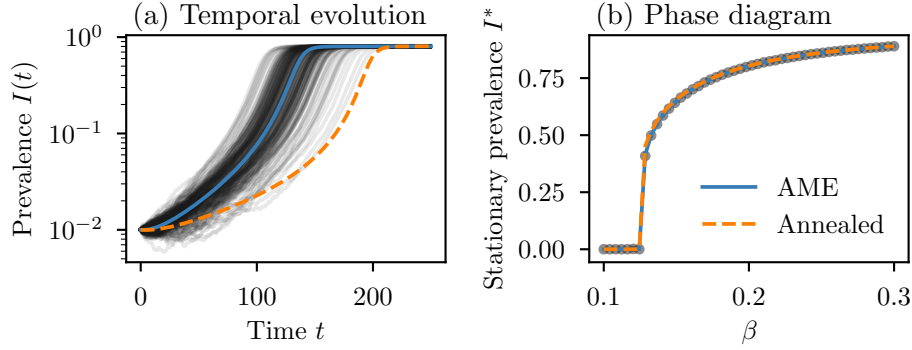

FIG. S7. Comparison with simulations for the temporal evolution and the phase diagram. We use an exponential dose distribution $\pi(\kappa; \lambda) \propto e^{-\kappa/\lambda}$, a power-law distribution of participation time $P(\tau) \propto \tau^{-\alpha-1}$ with $\alpha = 2$, a clearing window $\mathcal{T} \to \infty$, and $f(m) = 1$. We set the recovery probability to $\mu = 0.05$. We use Poisson distributions for both $\hat{P}(k)$ and $\hat{P}(m)$, with $\langle k \rangle = 5$ and $\langle m \rangle = 10$. For the simulations, we use random hypergraphs with $N = 10^4$ nodes. We use Eqs. (4)-(6) in the main text for the annealed solution and Eqs. (S61)-(S65) for the AMEs. (a) Temporal evolution. We performed 200 simulations each on a different hypergraph realization starting with a fraction $I(0) = 10^{-2}$ of infected nodes. The semitransparent solid black lines show the prevalence $I(t)$ for each realization. The AME solution is inside the region of high density of paths obtained from the simulations. (b) Phase diagram. For the simulations, we sampled the quasistationary state of the contagion process to estimate the stationary prevalence $I^*$. We performed measures on 10 hypergraph realizations. The semitransparent black circles correspond to the average prevalence and the vertical bars (too small to see) correspond to the standard deviation. Because we sample the quasistationary state, only the upper branch of the hysteresis loop is obtained. Both the annealed and AME solutions fit well with the simulations.

a randomly chosen state in the history with the current state of the system. At any time during the simulation, if the absorbing state is reached, we replace the current state with one chosen randomly from the history.

To estimate quantities in the stationary state, we first let the system thermalize by performing $10^3$ time steps. During this period, we use a short decorrelation period $t_d = 1$ to make sure the whole history has been updated since the initial state. Then, we perform an additional $10^3$ time steps during which we measure the system at regular intervals $t_d = 10$.

### C. Comparison with analytical predictions

In Fig. S7(a), we present multiple realization of the stochastic process (semitransparent solid black lines) for the temporal evolution of the prevalence $I(t)$. The darker region corresponds to a high density of paths for $I(t)$. Interestingly, while both analytical approaches have a similar qualitative temporal evolution, AMEs (solid blue line) best capture the temporal evolution, with their prediction being in the region of high density of paths obtained from the simulations. Therefore, dynamical correlations do matter here. Moreover, since $\alpha = 2$, we have a supralinear infection kernel ($\nu = 2$) leading to superexponential spread, a property validated by the simulations.

In Fig. S7(b), we compare the phase diagram obtained with our analytical approaches versus the simulations. In this case, both the annealed and the AMEs correctly predict the average prevalence in the stationary state—the prediction from AMEs are only slightly superior here. The simulations also confirm the discontinuous phase transition predicted by our analytical approaches.

### VI. CODE AND DATA

We provide Python modules to solve for the infection kernel, the invasion threshold $\beta_c$, the time evolution of the system, both for the annealed and quenched systems (AMEs), and to perform simulations of the contagion process. The code is available here [7].

In the main text, Fig. 1(a) was created using data from Refs. [8, 9]. They provide collections of COVID-19 superspreading events that were either reported in news outlet, on government websites, or in research papers.

We kept those where the attack rate, the duration of the event, and the size of the event (number of people attending) were either reported or could be calculated. The curated data is available here [10].

---

[1] S. Nadarajah, Exact distribution of the product of m gamma and n Pareto random variables, J. Comput. Appl. Math. **235**, 4496 (2011).




\begin{thebibliography}{70}%
\makeatletter
\providecommand \@ifxundefined [1]{%
 \@ifx{#1\undefined}
}%
\providecommand \@ifnum [1]{%
 \ifnum #1\expandafter \@firstoftwo
 \else \expandafter \@secondoftwo
 \fi
}%
\providecommand \@ifx [1]{%
 \ifx #1\expandafter \@firstoftwo
 \else \expandafter \@secondoftwo
 \fi
}%
\providecommand \natexlab [1]{#1}%
\providecommand \enquote  [1]{``#1''}%
\providecommand \bibnamefont  [1]{#1}%
\providecommand \bibfnamefont [1]{#1}%
\providecommand \citenamefont [1]{#1}%
\providecommand \href@noop [0]{\@secondoftwo}%
\providecommand \href [0]{\begingroup \@sanitize@url \@href}%
\providecommand \@href[1]{\@@startlink{#1}\@@href}%
\providecommand \@@href[1]{\endgroup#1\@@endlink}%
\providecommand \@sanitize@url [0]{\catcode `\\12\catcode `\$12\catcode
  `\&12\catcode `\#12\catcode `\^12\catcode `\_12\catcode `\%12\relax}%
\providecommand \@@startlink[1]{}%
\providecommand \@@endlink[0]{}%
\providecommand \url  [0]{\begingroup\@sanitize@url \@url }%
\providecommand \@url [1]{\endgroup\@href {#1}{\urlprefix }}%
\providecommand \urlprefix  [0]{URL }%
\providecommand \Eprint [0]{\href }%
\providecommand \doibase [0]{https://doi.org/}%
\providecommand \selectlanguage [0]{\@gobble}%
\providecommand \bibinfo  [0]{\@secondoftwo}%
\providecommand \bibfield  [0]{\@secondoftwo}%
\providecommand \translation [1]{[#1]}%
\providecommand \BibitemOpen [0]{}%
\providecommand \bibitemStop [0]{}%
\providecommand \bibitemNoStop [0]{.\EOS\space}%
\providecommand \EOS [0]{\spacefactor3000\relax}%
\providecommand \BibitemShut  [1]{\csname bibitem#1\endcsname}%
\let\auto@bib@innerbib\@empty
\bibitem [{\citenamefont {Rivers}\ \emph {et~al.}(2019)\citenamefont {Rivers},
  \citenamefont {Chretien}, \citenamefont {Riley}, \citenamefont {Pavlin},
  \citenamefont {Woodward}, \citenamefont {{Brett-Major}}, \citenamefont
  {Maljkovic~Berry}, \citenamefont {Morton}, \citenamefont {Jarman},
  \citenamefont {Biggerstaff}, \citenamefont {Johansson}, \citenamefont
  {Reich}, \citenamefont {Meyer}, \citenamefont {Snyder},\ and\ \citenamefont
  {Pollett}}]{rivers2019using}%
  \BibitemOpen
  \bibfield  {author} {\bibinfo {author} {\bibfnamefont {C.}~\bibnamefont
  {Rivers}}, \bibinfo {author} {\bibfnamefont {J.-P.}\ \bibnamefont
  {Chretien}}, \bibinfo {author} {\bibfnamefont {S.}~\bibnamefont {Riley}},
  \bibinfo {author} {\bibfnamefont {J.~A.}\ \bibnamefont {Pavlin}}, \bibinfo
  {author} {\bibfnamefont {A.}~\bibnamefont {Woodward}}, \bibinfo {author}
  {\bibfnamefont {D.}~\bibnamefont {{Brett-Major}}}, \bibinfo {author}
  {\bibfnamefont {I.}~\bibnamefont {Maljkovic~Berry}}, \bibinfo {author}
  {\bibfnamefont {L.}~\bibnamefont {Morton}}, \bibinfo {author} {\bibfnamefont
  {R.~G.}\ \bibnamefont {Jarman}}, \bibinfo {author} {\bibfnamefont
  {M.}~\bibnamefont {Biggerstaff}}, \bibinfo {author} {\bibfnamefont {M.~A.}\
  \bibnamefont {Johansson}}, \bibinfo {author} {\bibfnamefont {N.~G.}\
  \bibnamefont {Reich}}, \bibinfo {author} {\bibfnamefont {D.}~\bibnamefont
  {Meyer}}, \bibinfo {author} {\bibfnamefont {M.~R.}\ \bibnamefont {Snyder}},\
  and\ \bibinfo {author} {\bibfnamefont {S.}~\bibnamefont {Pollett}},\
  }\bibfield  {title} {\bibinfo {title} {Using ``outbreak science'' to
  strengthen the use of models during epidemics},\ }\href
  {https://doi.org/10.1038/s41467-019-11067-2} {\bibfield  {journal} {\bibinfo
  {journal} {Nat. Commun.}\ }\textbf {\bibinfo {volume} {10}},\ \bibinfo
  {pages} {3102} (\bibinfo {year} {2019})}\BibitemShut {NoStop}%
\bibitem [{\citenamefont {Epstein}(2008)}]{epstein2008why}%
  \BibitemOpen
  \bibfield  {author} {\bibinfo {author} {\bibfnamefont {J.~M.}\ \bibnamefont
  {Epstein}},\ }\bibfield  {title} {\bibinfo {title} {Why {{Model}}?},\ }\href
  {http://jasss.soc.surrey.ac.uk/11/4/12.html} {\bibfield  {journal} {\bibinfo
  {journal} {J. Artif. Soc. Soc. Simul.}\ }\textbf {\bibinfo {volume} {11}},\
  \bibinfo {pages} {12} (\bibinfo {year} {2008})}\BibitemShut {NoStop}%
\bibitem [{\citenamefont {Biggerstaff}\ \emph {et~al.}(2018)\citenamefont
  {Biggerstaff}, \citenamefont {Johansson}, \citenamefont {Alper},
  \citenamefont {Brooks}, \citenamefont {Chakraborty}, \citenamefont {Farrow},
  \citenamefont {Hyun}, \citenamefont {Kandula}, \citenamefont {McGowan},
  \citenamefont {Ramakrishnan}, \citenamefont {Rosenfeld}, \citenamefont
  {Shaman}, \citenamefont {Tibshirani}, \citenamefont {Tibshirani},
  \citenamefont {Vespignani}, \citenamefont {Yang}, \citenamefont {Zhang},\
  and\ \citenamefont {Reed}}]{biggerstaff2018results}%
  \BibitemOpen
  \bibfield  {author} {\bibinfo {author} {\bibfnamefont {M.}~\bibnamefont
  {Biggerstaff}}, \bibinfo {author} {\bibfnamefont {M.}~\bibnamefont
  {Johansson}}, \bibinfo {author} {\bibfnamefont {D.}~\bibnamefont {Alper}},
  \bibinfo {author} {\bibfnamefont {L.~C.}\ \bibnamefont {Brooks}}, \bibinfo
  {author} {\bibfnamefont {P.}~\bibnamefont {Chakraborty}}, \bibinfo {author}
  {\bibfnamefont {D.~C.}\ \bibnamefont {Farrow}}, \bibinfo {author}
  {\bibfnamefont {S.}~\bibnamefont {Hyun}}, \bibinfo {author} {\bibfnamefont
  {S.}~\bibnamefont {Kandula}}, \bibinfo {author} {\bibfnamefont
  {C.}~\bibnamefont {McGowan}}, \bibinfo {author} {\bibfnamefont
  {N.}~\bibnamefont {Ramakrishnan}}, \bibinfo {author} {\bibfnamefont
  {R.}~\bibnamefont {Rosenfeld}}, \bibinfo {author} {\bibfnamefont
  {J.}~\bibnamefont {Shaman}}, \bibinfo {author} {\bibfnamefont
  {R.}~\bibnamefont {Tibshirani}}, \bibinfo {author} {\bibfnamefont {R.~J.}\
  \bibnamefont {Tibshirani}}, \bibinfo {author} {\bibfnamefont
  {A.}~\bibnamefont {Vespignani}}, \bibinfo {author} {\bibfnamefont
  {W.}~\bibnamefont {Yang}}, \bibinfo {author} {\bibfnamefont {Q.}~\bibnamefont
  {Zhang}},\ and\ \bibinfo {author} {\bibfnamefont {C.}~\bibnamefont {Reed}},\
  }\bibfield  {title} {\bibinfo {title} {Results from the second year of a
  collaborative effort to forecast influenza seasons in the {{United
  States}}},\ }\href {https://doi.org/10.1016/j.epidem.2018.02.003} {\bibfield
  {journal} {\bibinfo  {journal} {Epidemics}\ }\textbf {\bibinfo {volume}
  {24}},\ \bibinfo {pages} {26} (\bibinfo {year} {2018})}\BibitemShut {NoStop}%
\bibitem [{\citenamefont {Liu}\ \emph {et~al.}(1986)\citenamefont {Liu},
  \citenamefont {Levin},\ and\ \citenamefont {Iwasa}}]{liu1986influence}%
  \BibitemOpen
  \bibfield  {author} {\bibinfo {author} {\bibfnamefont {W.-m.}\ \bibnamefont
  {Liu}}, \bibinfo {author} {\bibfnamefont {S.~A.}\ \bibnamefont {Levin}},\
  and\ \bibinfo {author} {\bibfnamefont {Y.}~\bibnamefont {Iwasa}},\ }\bibfield
   {title} {\bibinfo {title} {Influence of nonlinear incidence rates upon the
  behavior of {{SIRS}} epidemiological models},\ }\href
  {https://doi.org/10.1007/BF00276956} {\bibfield  {journal} {\bibinfo
  {journal} {J. Math. Biol.}\ }\textbf {\bibinfo {volume} {23}},\ \bibinfo
  {pages} {187} (\bibinfo {year} {1986})}\BibitemShut {NoStop}%
\bibitem [{\citenamefont {Hethcote}\ and\ \citenamefont {{van den
  Driessche}}(1991)}]{hethcote1991some}%
  \BibitemOpen
  \bibfield  {author} {\bibinfo {author} {\bibfnamefont {H.~W.}\ \bibnamefont
  {Hethcote}}\ and\ \bibinfo {author} {\bibfnamefont {P.}~\bibnamefont {{van
  den Driessche}}},\ }\bibfield  {title} {\bibinfo {title} {Some
  epidemiological models with nonlinear incidence},\ }\href
  {https://doi.org/10.1007/BF00160539} {\bibfield  {journal} {\bibinfo
  {journal} {J. Math. Biol.}\ }\textbf {\bibinfo {volume} {29}},\ \bibinfo
  {pages} {271} (\bibinfo {year} {1991})}\BibitemShut {NoStop}%
\bibitem [{\citenamefont {Centola}(2018)}]{centola2018how}%
  \BibitemOpen
  \bibfield  {author} {\bibinfo {author} {\bibfnamefont {D.}~\bibnamefont
  {Centola}},\ }\href@noop {} {\emph {\bibinfo {title} {How Behavior Spreads:
  The Science of Complex Contagions}}}\ (\bibinfo  {publisher} {{Princeton
  University Press}},\ \bibinfo {year} {2018})\BibitemShut {NoStop}%
\bibitem [{\citenamefont {Lehmann}\ and\ \citenamefont
  {Ahn}(2018)}]{Lehmann2018}%
  \BibitemOpen
  \bibfield  {author} {\bibinfo {author} {\bibfnamefont {S.}~\bibnamefont
  {Lehmann}}\ and\ \bibinfo {author} {\bibfnamefont {Y.-.}\ \bibnamefont
  {Ahn}},\ }\href {https://doi.org/10.1007/978-3-319-77332-2} {\emph {\bibinfo
  {title} {{Complex Spreading Phenomena in Social Systems}}}}\ (\bibinfo
  {publisher} {Springer},\ \bibinfo {year} {2018})\BibitemShut {NoStop}%
\bibitem [{\citenamefont {Centola}(2010)}]{centola2010spread}%
  \BibitemOpen
  \bibfield  {author} {\bibinfo {author} {\bibfnamefont {D.}~\bibnamefont
  {Centola}},\ }\bibfield  {title} {\bibinfo {title} {The {{Spread}} of
  {{Behavior}} in an {{Online Social Network Experiment}}},\ }\href
  {https://doi.org/10.1126/science.1185231} {\bibfield  {journal} {\bibinfo
  {journal} {Science}\ }\textbf {\bibinfo {volume} {329}},\ \bibinfo {pages}
  {1194} (\bibinfo {year} {2010})}\BibitemShut {NoStop}%
\bibitem [{\citenamefont {Busenberg}\ \emph {et~al.}(1991)\citenamefont
  {Busenberg}, \citenamefont {Iannelli},\ and\ \citenamefont
  {Thieme}}]{busenberg1991global}%
  \BibitemOpen
  \bibfield  {author} {\bibinfo {author} {\bibfnamefont {S.~N.}\ \bibnamefont
  {Busenberg}}, \bibinfo {author} {\bibfnamefont {M.}~\bibnamefont
  {Iannelli}},\ and\ \bibinfo {author} {\bibfnamefont {H.~R.}\ \bibnamefont
  {Thieme}},\ }\bibfield  {title} {\bibinfo {title} {Global {{Behavior}} of an
  {{Age}}-{{Structured Epidemic Model}}},\ }\href
  {https://doi.org/10.1137/0522069} {\bibfield  {journal} {\bibinfo  {journal}
  {SIAM J. Math. Anal.}\ }\textbf {\bibinfo {volume} {22}},\ \bibinfo {pages}
  {1065} (\bibinfo {year} {1991})}\BibitemShut {NoStop}%
\bibitem [{\citenamefont {Feng}\ and\ \citenamefont
  {Thieme}(2000)}]{feng2000endemic}%
  \BibitemOpen
  \bibfield  {author} {\bibinfo {author} {\bibfnamefont {Z.}~\bibnamefont
  {Feng}}\ and\ \bibinfo {author} {\bibfnamefont {H.~R.}\ \bibnamefont
  {Thieme}},\ }\bibfield  {title} {\bibinfo {title} {Endemic {{Models}} with
  {{Arbitrarily Distributed Periods}} of {{Infection I}}: {{Fundamental
  Properties}} of the {{Model}}},\ }\href
  {https://doi.org/10.1137/S0036139998347834} {\bibfield  {journal} {\bibinfo
  {journal} {SIAM J. Appl. Math.}\ }\textbf {\bibinfo {volume} {61}},\ \bibinfo
  {pages} {803} (\bibinfo {year} {2000})}\BibitemShut {NoStop}%
\bibitem [{\citenamefont {Hethcote}\ and\ \citenamefont
  {Van~Ark}(1987)}]{hethcote1987epidemiological}%
  \BibitemOpen
  \bibfield  {author} {\bibinfo {author} {\bibfnamefont {H.~W.}\ \bibnamefont
  {Hethcote}}\ and\ \bibinfo {author} {\bibfnamefont {J.~W.}\ \bibnamefont
  {Van~Ark}},\ }\bibfield  {title} {\bibinfo {title} {Epidemiological models
  for heterogeneous populations: Proportionate mixing, parameter estimation,
  and immunization programs},\ }\href
  {https://doi.org/10.1016/0025-5564(87)90044-7} {\bibfield  {journal}
  {\bibinfo  {journal} {Math. Biosci.}\ }\textbf {\bibinfo {volume} {84}},\
  \bibinfo {pages} {85} (\bibinfo {year} {1987})}\BibitemShut {NoStop}%
\bibitem [{\citenamefont {Hethcote}(1996)}]{hethcote1996modeling}%
  \BibitemOpen
  \bibfield  {author} {\bibinfo {author} {\bibfnamefont {H.~W.}\ \bibnamefont
  {Hethcote}},\ }\bibfield  {title} {\bibinfo {title} {Modeling heterogeneous
  mixing in infectious disease dynamics},\ }in\ \href
  {https://doi.org/10.1017/CBO9780511662935.030} {\emph {\bibinfo {booktitle}
  {Models for {{Infectious Human Diseases}}: {{Their Structure}} and
  {{Relation}} to {{Data}}}}}\ (\bibinfo  {publisher} {{Cambridge University
  Press}},\ \bibinfo {year} {1996})\ \bibinfo {edition} {1st}\ ed.,\ pp.\
  \bibinfo {pages} {215--238}\BibitemShut {NoStop}%
\bibitem [{\citenamefont {{Pastor-Satorras}}\ and\ \citenamefont
  {Vespignani}(2001)}]{pastor-satorras2001epidemic}%
  \BibitemOpen
  \bibfield  {author} {\bibinfo {author} {\bibfnamefont {R.}~\bibnamefont
  {{Pastor-Satorras}}}\ and\ \bibinfo {author} {\bibfnamefont {A.}~\bibnamefont
  {Vespignani}},\ }\bibfield  {title} {\bibinfo {title} {Epidemic {{Spreading}}
  in {{Scale}}-{{Free Networks}}},\ }\href
  {https://doi.org/10.1103/PhysRevLett.86.3200} {\bibfield  {journal} {\bibinfo
   {journal} {Phys. Rev. Lett.}\ }\textbf {\bibinfo {volume} {86}},\ \bibinfo
  {pages} {3200} (\bibinfo {year} {2001})}\BibitemShut {NoStop}%
\bibitem [{\citenamefont {{Pastor-Satorras}}\ \emph {et~al.}(2015)\citenamefont
  {{Pastor-Satorras}}, \citenamefont {Castellano}, \citenamefont
  {Van~Mieghem},\ and\ \citenamefont
  {Vespignani}}]{pastor-satorras2015epidemic}%
  \BibitemOpen
  \bibfield  {author} {\bibinfo {author} {\bibfnamefont {R.}~\bibnamefont
  {{Pastor-Satorras}}}, \bibinfo {author} {\bibfnamefont {C.}~\bibnamefont
  {Castellano}}, \bibinfo {author} {\bibfnamefont {P.}~\bibnamefont
  {Van~Mieghem}},\ and\ \bibinfo {author} {\bibfnamefont {A.}~\bibnamefont
  {Vespignani}},\ }\bibfield  {title} {\bibinfo {title} {Epidemic processes in
  complex networks},\ }\href {https://doi.org/10.1103/RevModPhys.87.925}
  {\bibfield  {journal} {\bibinfo  {journal} {Rev. Mod. Phys.}\ }\textbf
  {\bibinfo {volume} {87}},\ \bibinfo {pages} {925} (\bibinfo {year}
  {2015})}\BibitemShut {NoStop}%
\bibitem [{\citenamefont {Swinkels}(2020)}]{SSdata_original}%
  \BibitemOpen
  \bibfield  {author} {\bibinfo {author} {\bibfnamefont {K.}~\bibnamefont
  {Swinkels}},\ }\href {http://www.superspreadingdatabase.com/} {\bibinfo
  {title} {{SARS-CoV-2} superspreading events from around the world}} (\bibinfo
  {year} {2020}),\ \bibinfo {note} {accessed: 2020-07-02}\BibitemShut {NoStop}%
\bibitem [{\citenamefont {Leclerc}\ \emph {et~al.}(2020)\citenamefont
  {Leclerc}, \citenamefont {Fuller}, \citenamefont {Knight}, \citenamefont
  {Funk}, \citenamefont {Knight},\ and\ \citenamefont {Group}}]{leclerc2020}%
  \BibitemOpen
  \bibfield  {author} {\bibinfo {author} {\bibfnamefont {Q.~J.}\ \bibnamefont
  {Leclerc}}, \bibinfo {author} {\bibfnamefont {N.~M.}\ \bibnamefont {Fuller}},
  \bibinfo {author} {\bibfnamefont {L.~E.}\ \bibnamefont {Knight}}, \bibinfo
  {author} {\bibfnamefont {S.}~\bibnamefont {Funk}}, \bibinfo {author}
  {\bibfnamefont {G.~M.}\ \bibnamefont {Knight}},\ and\ \bibinfo {author}
  {\bibfnamefont {C.~C.-.~W.}\ \bibnamefont {Group}},\ }\bibfield  {title}
  {\bibinfo {title} {What settings have been linked to sars-cov-2 transmission
  clusters?},\ }\href {https://doi.org/10.12688/wellcomeopenres.15889.2}
  {\bibfield  {journal} {\bibinfo  {journal} {Wellcome Open Research}\ }\textbf
  {\bibinfo {volume} {5}} (\bibinfo {year} {2020})}\BibitemShut {NoStop}%
\bibitem [{SM()}]{SM}%
  \BibitemOpen
  \href@noop {} {}\bibinfo {note} {See Supplemental Material for a broader
  derivation of the infection kernel, the consideration of other sources of
  heterogeneity, the characterization of the phase transition, and a treatment
  of dynamical correlations, which includes
  Refs.~\cite{nadarajah2011exact,st-onge2021social,st-onge2021master,fosdick2018configuring,chodrow2020configuration,deoliveira2005,SSdata_original,leclerc2020,SSdata_curated,github-repository}.}\BibitemShut
  {Stop}%
\bibitem [{fre()}]{freepik}%
  \BibitemOpen
  \href@noop {} {}\bibinfo {note} {Icons made by Freepik from
  www.flaticon.com}\BibitemShut {NoStop}%
\bibitem [{\citenamefont {Balasubramanian}\ \emph {et~al.}(1994)\citenamefont
  {Balasubramanian}, \citenamefont {Wiegeshaus}, \citenamefont {Taylor},\ and\
  \citenamefont {Smith}}]{balasubramanian1994pathogenesis}%
  \BibitemOpen
  \bibfield  {author} {\bibinfo {author} {\bibfnamefont {V.}~\bibnamefont
  {Balasubramanian}}, \bibinfo {author} {\bibfnamefont {E.~H.}\ \bibnamefont
  {Wiegeshaus}}, \bibinfo {author} {\bibfnamefont {B.~T.}\ \bibnamefont
  {Taylor}},\ and\ \bibinfo {author} {\bibfnamefont {D.~W.}\ \bibnamefont
  {Smith}},\ }\bibfield  {title} {\bibinfo {title} {Pathogenesis of
  tuberculosis: Pathway to apical localization},\ }\href
  {https://doi.org/10.1016/0962-8479(94)90002-7} {\bibfield  {journal}
  {\bibinfo  {journal} {Tuber. Lung Dis.}\ }\textbf {\bibinfo {volume} {75}},\
  \bibinfo {pages} {168} (\bibinfo {year} {1994})}\BibitemShut {NoStop}%
\bibitem [{\citenamefont {LaRocque}\ and\ \citenamefont
  {Calderwood}(2015)}]{larocque2015syndromes}%
  \BibitemOpen
  \bibfield  {author} {\bibinfo {author} {\bibfnamefont {R.~C.}\ \bibnamefont
  {LaRocque}}\ and\ \bibinfo {author} {\bibfnamefont {S.~B.}\ \bibnamefont
  {Calderwood}},\ }\bibfield  {title} {\bibinfo {title} {Syndromes of enteric
  infection},\ }in\ \href@noop {} {\emph {\bibinfo {booktitle} {Mandell,
  Douglas, and Bennett's Principles and Practice of Infectious Diseases}}}\
  (\bibinfo  {publisher} {Elsevier},\ \bibinfo {year} {2015})\ pp.\ \bibinfo
  {pages} {1238--1247}\BibitemShut {NoStop}%
\bibitem [{\citenamefont {Weber}\ and\ \citenamefont
  {Stilianakis}(2008)}]{weber2008inactivation}%
  \BibitemOpen
  \bibfield  {author} {\bibinfo {author} {\bibfnamefont {T.~P.}\ \bibnamefont
  {Weber}}\ and\ \bibinfo {author} {\bibfnamefont {N.~I.}\ \bibnamefont
  {Stilianakis}},\ }\bibfield  {title} {\bibinfo {title} {Inactivation of
  influenza {{A}} viruses in the environment and modes of transmission: {{A}}
  critical review},\ }\href {https://doi.org/10.1016/j.jinf.2008.08.013}
  {\bibfield  {journal} {\bibinfo  {journal} {J. Infect.}\ }\textbf {\bibinfo
  {volume} {57}},\ \bibinfo {pages} {361} (\bibinfo {year} {2008})}\BibitemShut
  {NoStop}%
\bibitem [{\citenamefont {Gama}\ \emph {et~al.}(2012)\citenamefont {Gama},
  \citenamefont {Abby}, \citenamefont {Vieira-Silva}, \citenamefont
  {Dionisio},\ and\ \citenamefont {Rocha}}]{gama2012immune}%
  \BibitemOpen
  \bibfield  {author} {\bibinfo {author} {\bibfnamefont {J.~A.}\ \bibnamefont
  {Gama}}, \bibinfo {author} {\bibfnamefont {S.~S.}\ \bibnamefont {Abby}},
  \bibinfo {author} {\bibfnamefont {S.}~\bibnamefont {Vieira-Silva}}, \bibinfo
  {author} {\bibfnamefont {F.}~\bibnamefont {Dionisio}},\ and\ \bibinfo
  {author} {\bibfnamefont {E.~P.~C.}\ \bibnamefont {Rocha}},\ }\bibfield
  {title} {\bibinfo {title} {Immune subversion and quorum-sensing shape the
  variation in infectious dose among bacterial pathogens},\ }\href
  {https://doi.org/10.1371/journal.ppat.1002503} {\bibfield  {journal}
  {\bibinfo  {journal} {PLOS Pathog.}\ }\textbf {\bibinfo {volume} {8}},\
  \bibinfo {pages} {e1002503} (\bibinfo {year} {2012})}\BibitemShut {NoStop}%
\bibitem [{\citenamefont {Finlay}\ and\ \citenamefont
  {Falkow}(1997)}]{finlay1997common}%
  \BibitemOpen
  \bibfield  {author} {\bibinfo {author} {\bibfnamefont {B.~B.}\ \bibnamefont
  {Finlay}}\ and\ \bibinfo {author} {\bibfnamefont {S.}~\bibnamefont
  {Falkow}},\ }\bibfield  {title} {\bibinfo {title} {Common themes in microbial
  pathogenicity revisited.},\ }\href {https://mmbr.asm.org/content/61/2/136}
  {\bibfield  {journal} {\bibinfo  {journal} {Microbiol. Mol. Biol. Rev.}\
  }\textbf {\bibinfo {volume} {61}},\ \bibinfo {pages} {136} (\bibinfo {year}
  {1997})}\BibitemShut {NoStop}%
\bibitem [{\citenamefont {Hornef}\ \emph {et~al.}(2002)\citenamefont {Hornef},
  \citenamefont {Wick}, \citenamefont {Rhen},\ and\ \citenamefont
  {Normark}}]{hornef2002bacterial}%
  \BibitemOpen
  \bibfield  {author} {\bibinfo {author} {\bibfnamefont {M.~W.}\ \bibnamefont
  {Hornef}}, \bibinfo {author} {\bibfnamefont {M.~J.}\ \bibnamefont {Wick}},
  \bibinfo {author} {\bibfnamefont {M.}~\bibnamefont {Rhen}},\ and\ \bibinfo
  {author} {\bibfnamefont {S.}~\bibnamefont {Normark}},\ }\bibfield  {title}
  {\bibinfo {title} {Bacterial strategies for overcoming host innate and
  adaptive immune responses},\ }\href {https://doi.org/10.1038/ni1102-1033}
  {\bibfield  {journal} {\bibinfo  {journal} {Nat. immunol.}\ }\textbf
  {\bibinfo {volume} {3}},\ \bibinfo {pages} {1033} (\bibinfo {year}
  {2002})}\BibitemShut {NoStop}%
\bibitem [{\citenamefont {Lipsitch}\ and\ \citenamefont
  {O'Hagan}(2007)}]{lipsitch2007patterns}%
  \BibitemOpen
  \bibfield  {author} {\bibinfo {author} {\bibfnamefont {M.}~\bibnamefont
  {Lipsitch}}\ and\ \bibinfo {author} {\bibfnamefont {J.~J.}\ \bibnamefont
  {O'Hagan}},\ }\bibfield  {title} {\bibinfo {title} {Patterns of antigenic
  diversity and the mechanisms that maintain them},\ }\href
  {https://doi.org/10.1098/rsif.2007.0229} {\bibfield  {journal} {\bibinfo
  {journal} {J. R. Soc. Interface}\ }\textbf {\bibinfo {volume} {4}},\ \bibinfo
  {pages} {787} (\bibinfo {year} {2007})}\BibitemShut {NoStop}%
\bibitem [{\citenamefont {Casadevall}(2008)}]{casadevall2008evolution}%
  \BibitemOpen
  \bibfield  {author} {\bibinfo {author} {\bibfnamefont {A.}~\bibnamefont
  {Casadevall}},\ }\bibfield  {title} {\bibinfo {title} {Evolution of
  intracellular pathogens},\ }\href
  {https://doi.org/10.1146/annurev.micro.61.080706.093305} {\bibfield
  {journal} {\bibinfo  {journal} {Annu. Rev. Microbiol.}\ }\textbf {\bibinfo
  {volume} {62}},\ \bibinfo {pages} {19} (\bibinfo {year} {2008})}\BibitemShut
  {NoStop}%
\bibitem [{\citenamefont {Du}\ and\ \citenamefont
  {Yuan}(2020)}]{du2020mathematical}%
  \BibitemOpen
  \bibfield  {author} {\bibinfo {author} {\bibfnamefont {S.~Q.}\ \bibnamefont
  {Du}}\ and\ \bibinfo {author} {\bibfnamefont {W.}~\bibnamefont {Yuan}},\
  }\bibfield  {title} {\bibinfo {title} {Mathematical modeling of interaction
  between innate and adaptive immune responses in {COVID-19} and implications
  for viral pathogenesis},\ }\href {https://doi.org/10.1002/jmv.25866}
  {\bibfield  {journal} {\bibinfo  {journal} {J. Med. Virol.}\ }\textbf
  {\bibinfo {volume} {92}},\ \bibinfo {pages} {1615} (\bibinfo {year}
  {2020})}\BibitemShut {NoStop}%
\bibitem [{\citenamefont {Beauchemin}\ and\ \citenamefont
  {Handel}(2011)}]{beauchemin2011review}%
  \BibitemOpen
  \bibfield  {author} {\bibinfo {author} {\bibfnamefont {C.~A.}\ \bibnamefont
  {Beauchemin}}\ and\ \bibinfo {author} {\bibfnamefont {A.}~\bibnamefont
  {Handel}},\ }\bibfield  {title} {\bibinfo {title} {{A review of mathematical
  models of influenza A infections within a host or cell culture: lessons
  learned and challenges ahead}},\ }\href
  {https://doi.org/10.1186/1471-2458-11-S1-S7} {\bibfield  {journal} {\bibinfo
  {journal} {BMC Public Health}\ }\textbf {\bibinfo {volume} {11}},\ \bibinfo
  {pages} {S7} (\bibinfo {year} {2011})}\BibitemShut {NoStop}%
\bibitem [{\citenamefont {Liu}\ \emph {et~al.}(2020)\citenamefont {Liu},
  \citenamefont {Eggo},\ and\ \citenamefont {Kucharski}}]{liu2020secondary}%
  \BibitemOpen
  \bibfield  {author} {\bibinfo {author} {\bibfnamefont {Y.}~\bibnamefont
  {Liu}}, \bibinfo {author} {\bibfnamefont {R.~M.}\ \bibnamefont {Eggo}},\ and\
  \bibinfo {author} {\bibfnamefont {A.~J.}\ \bibnamefont {Kucharski}},\
  }\bibfield  {title} {\bibinfo {title} {Secondary attack rate and
  superspreading events for sars-cov-2},\ }\href
  {https://doi.org/https://doi.org/10.1016/S0140-6736(20)30462-1} {\bibfield
  {journal} {\bibinfo  {journal} {Lancet}\ }\textbf {\bibinfo {volume} {395}},\
  \bibinfo {pages} {e47} (\bibinfo {year} {2020})}\BibitemShut {NoStop}%
\bibitem [{\citenamefont {Wong}\ and\ \citenamefont
  {Collins}(2020)}]{wong2020evidence}%
  \BibitemOpen
  \bibfield  {author} {\bibinfo {author} {\bibfnamefont {F.}~\bibnamefont
  {Wong}}\ and\ \bibinfo {author} {\bibfnamefont {J.~J.}\ \bibnamefont
  {Collins}},\ }\bibfield  {title} {\bibinfo {title} {Evidence that coronavirus
  superspreading is fat-tailed},\ }\href
  {https://doi.org/10.1073/pnas.2018490117} {\bibfield  {journal} {\bibinfo
  {journal} {Proc. Natl. Acad. Sci. U.S.A.}\ }\textbf {\bibinfo {volume}
  {117}},\ \bibinfo {pages} {29416} (\bibinfo {year} {2020})}\BibitemShut
  {NoStop}%
\bibitem [{\citenamefont {Althouse}\ \emph {et~al.}(2020)\citenamefont
  {Althouse}, \citenamefont {Wenger}, \citenamefont {Miller}, \citenamefont
  {Scarpino}, \citenamefont {Allard}, \citenamefont {{H{\'e}bert-Dufresne}},\
  and\ \citenamefont {Hu}}]{althouse2020superspreading}%
  \BibitemOpen
  \bibfield  {author} {\bibinfo {author} {\bibfnamefont {B.~M.}\ \bibnamefont
  {Althouse}}, \bibinfo {author} {\bibfnamefont {E.~A.}\ \bibnamefont
  {Wenger}}, \bibinfo {author} {\bibfnamefont {J.~C.}\ \bibnamefont {Miller}},
  \bibinfo {author} {\bibfnamefont {S.~V.}\ \bibnamefont {Scarpino}}, \bibinfo
  {author} {\bibfnamefont {A.}~\bibnamefont {Allard}}, \bibinfo {author}
  {\bibfnamefont {L.}~\bibnamefont {{H{\'e}bert-Dufresne}}},\ and\ \bibinfo
  {author} {\bibfnamefont {H.}~\bibnamefont {Hu}},\ }\bibfield  {title}
  {\bibinfo {title} {Superspreading events in the transmission dynamics of
  {{SARS}}-{{CoV}}-2: {{Opportunities}} for interventions and control},\ }\href
  {https://doi.org/10.1371/journal.pbio.3000897} {\bibfield  {journal}
  {\bibinfo  {journal} {PLOS Biol.}\ }\textbf {\bibinfo {volume} {18}},\
  \bibinfo {pages} {e3000897} (\bibinfo {year} {2020})}\BibitemShut {NoStop}%
\bibitem [{\citenamefont {Endo}\ \emph {et~al.}(2020)\citenamefont {Endo},
  \citenamefont {Abbott}, \citenamefont {Kucharski},\ and\ \citenamefont
  {Funk}}]{endo2020estimating}%
  \BibitemOpen
  \bibfield  {author} {\bibinfo {author} {\bibfnamefont {A.}~\bibnamefont
  {Endo}}, \bibinfo {author} {\bibfnamefont {S.}~\bibnamefont {Abbott}},
  \bibinfo {author} {\bibfnamefont {A.~J.}\ \bibnamefont {Kucharski}},\ and\
  \bibinfo {author} {\bibfnamefont {S.}~\bibnamefont {Funk}},\ }\bibfield
  {title} {\bibinfo {title} {Estimating the overdispersion in {COVID-19}
  transmission using outbreak sizes outside china},\ }\href
  {https://doi.org/10.12688/wellcomeopenres.15842.3} {\bibfield  {journal}
  {\bibinfo  {journal} {Wellcome Open Res.}\ }\textbf {\bibinfo {volume} {5}},\
  \bibinfo {pages} {67} (\bibinfo {year} {2020})}\BibitemShut {NoStop}%
\bibitem [{\citenamefont {Bi}\ \emph {et~al.}(2020)\citenamefont {Bi} \emph
  {et~al.}}]{bi2020epidemiology}%
  \BibitemOpen
  \bibfield  {author} {\bibinfo {author} {\bibfnamefont {Q.}~\bibnamefont {Bi}}
  \emph {et~al.},\ }\bibfield  {title} {\bibinfo {title} {Epidemiology and
  transmission of {COVID-19} in 391 cases and 1286 of their close contacts in
  shenzhen, china: a retrospective cohort study},\ }\href
  {https://doi.org/10.1016/S1473-3099(20)30287-5} {\bibfield  {journal}
  {\bibinfo  {journal} {Lancet Infect. Dis.}\ }\textbf {\bibinfo {volume}
  {20}},\ \bibinfo {pages} {911} (\bibinfo {year} {2020})}\BibitemShut
  {NoStop}%
\bibitem [{\citenamefont {Miller}\ \emph {et~al.}(2020)\citenamefont {Miller},
  \citenamefont {Martin}, \citenamefont {Harel}, \citenamefont {Tirosh},
  \citenamefont {Kustin}, \citenamefont {Meir}, \citenamefont {Sorek},
  \citenamefont {Gefen-Halevi}, \citenamefont {Amit}, \citenamefont {Vorontsov}
  \emph {et~al.}}]{miller2020full}%
  \BibitemOpen
  \bibfield  {author} {\bibinfo {author} {\bibfnamefont {D.}~\bibnamefont
  {Miller}}, \bibinfo {author} {\bibfnamefont {M.~A.}\ \bibnamefont {Martin}},
  \bibinfo {author} {\bibfnamefont {N.}~\bibnamefont {Harel}}, \bibinfo
  {author} {\bibfnamefont {O.}~\bibnamefont {Tirosh}}, \bibinfo {author}
  {\bibfnamefont {T.}~\bibnamefont {Kustin}}, \bibinfo {author} {\bibfnamefont
  {M.}~\bibnamefont {Meir}}, \bibinfo {author} {\bibfnamefont {N.}~\bibnamefont
  {Sorek}}, \bibinfo {author} {\bibfnamefont {S.}~\bibnamefont {Gefen-Halevi}},
  \bibinfo {author} {\bibfnamefont {S.}~\bibnamefont {Amit}}, \bibinfo {author}
  {\bibfnamefont {O.}~\bibnamefont {Vorontsov}}, \emph {et~al.},\ }\bibfield
  {title} {\bibinfo {title} {Full genome viral sequences inform patterns of
  {SARS-CoV-2} spread into and within israel},\ }\href
  {https://doi.org/10.1038/s41467-020-19248-0} {\bibfield  {journal} {\bibinfo
  {journal} {Nat. Commun.}\ }\textbf {\bibinfo {volume} {11}},\ \bibinfo
  {pages} {5518} (\bibinfo {year} {2020})}\BibitemShut {NoStop}%
\bibitem [{\citenamefont {Lau}\ \emph {et~al.}(2020)\citenamefont {Lau},
  \citenamefont {Grenfell}, \citenamefont {Thomas}, \citenamefont {Bryan},
  \citenamefont {Nelson},\ and\ \citenamefont {Lopman}}]{lau2020charac}%
  \BibitemOpen
  \bibfield  {author} {\bibinfo {author} {\bibfnamefont {M.~S.~Y.}\
  \bibnamefont {Lau}}, \bibinfo {author} {\bibfnamefont {B.}~\bibnamefont
  {Grenfell}}, \bibinfo {author} {\bibfnamefont {M.}~\bibnamefont {Thomas}},
  \bibinfo {author} {\bibfnamefont {M.}~\bibnamefont {Bryan}}, \bibinfo
  {author} {\bibfnamefont {K.}~\bibnamefont {Nelson}},\ and\ \bibinfo {author}
  {\bibfnamefont {B.}~\bibnamefont {Lopman}},\ }\bibfield  {title} {\bibinfo
  {title} {Characterizing superspreading events and age-specific infectiousness
  of {SARS-CoV-2} transmission in {Georgia, USA}},\ }\href
  {https://doi.org/10.1073/pnas.2011802117} {\bibfield  {journal} {\bibinfo
  {journal} {Proc. Natl. Acad. Sci. U.S.A.}\ }\textbf {\bibinfo {volume}
  {117}},\ \bibinfo {pages} {22430} (\bibinfo {year} {2020})},\ \Eprint
  {https://arxiv.org/abs/https://www.pnas.org/content/117/36/22430.full.pdf}
  {https://www.pnas.org/content/117/36/22430.full.pdf} \BibitemShut {NoStop}%
\bibitem [{\citenamefont {Nielsen}\ \emph {et~al.}(2021)\citenamefont
  {Nielsen}, \citenamefont {Simonsen},\ and\ \citenamefont
  {Sneppen}}]{nielsen2021covid}%
  \BibitemOpen
  \bibfield  {author} {\bibinfo {author} {\bibfnamefont {B.~F.}\ \bibnamefont
  {Nielsen}}, \bibinfo {author} {\bibfnamefont {L.}~\bibnamefont {Simonsen}},\
  and\ \bibinfo {author} {\bibfnamefont {K.}~\bibnamefont {Sneppen}},\
  }\bibfield  {title} {\bibinfo {title} {Covid-19 superspreading suggests
  mitigation by social network modulation},\ }\href
  {https://doi.org/10.1103/PhysRevLett.126.118301} {\bibfield  {journal}
  {\bibinfo  {journal} {Phys. Rev. Lett.}\ }\textbf {\bibinfo {volume} {126}},\
  \bibinfo {pages} {118301} (\bibinfo {year} {2021})}\BibitemShut {NoStop}%
\bibitem [{\citenamefont {Lloyd-Smith}\ \emph {et~al.}(2005)\citenamefont
  {Lloyd-Smith}, \citenamefont {Schreiber}, \citenamefont {Kopp},\ and\
  \citenamefont {Getz}}]{lloyd2005superspreading}%
  \BibitemOpen
  \bibfield  {author} {\bibinfo {author} {\bibfnamefont {J.~O.}\ \bibnamefont
  {Lloyd-Smith}}, \bibinfo {author} {\bibfnamefont {S.~J.}\ \bibnamefont
  {Schreiber}}, \bibinfo {author} {\bibfnamefont {P.~E.}\ \bibnamefont
  {Kopp}},\ and\ \bibinfo {author} {\bibfnamefont {W.~M.}\ \bibnamefont
  {Getz}},\ }\bibfield  {title} {\bibinfo {title} {Superspreading and the
  effect of individual variation on disease emergence},\ }\href
  {https://doi.org/10.1038/nature04153} {\bibfield  {journal} {\bibinfo
  {journal} {Nature}\ }\textbf {\bibinfo {volume} {438}},\ \bibinfo {pages}
  {355} (\bibinfo {year} {2005})}\BibitemShut {NoStop}%
\bibitem [{\citenamefont {Karsai}\ \emph {et~al.}(2018)\citenamefont {Karsai},
  \citenamefont {Jo},\ and\ \citenamefont {Kaski}}]{karsai2018bursty}%
  \BibitemOpen
  \bibfield  {author} {\bibinfo {author} {\bibfnamefont {M.}~\bibnamefont
  {Karsai}}, \bibinfo {author} {\bibfnamefont {H.-H.}\ \bibnamefont {Jo}},\
  and\ \bibinfo {author} {\bibfnamefont {K.}~\bibnamefont {Kaski}},\ }\href
  {https://doi.org/10.1007/978-3-319-68540-3} {\emph {\bibinfo {title} {Bursty
  {{Human Dynamics}}}}}\ (\bibinfo  {publisher} {{Springer International
  Publishing}},\ \bibinfo {year} {2018})\BibitemShut {NoStop}%
\bibitem [{\citenamefont {Karsai}\ \emph {et~al.}(2012)\citenamefont {Karsai},
  \citenamefont {Kaski}, \citenamefont {Barab{\'a}si},\ and\ \citenamefont
  {Kert{\'e}sz}}]{karsai2012universal}%
  \BibitemOpen
  \bibfield  {author} {\bibinfo {author} {\bibfnamefont {M.}~\bibnamefont
  {Karsai}}, \bibinfo {author} {\bibfnamefont {K.}~\bibnamefont {Kaski}},
  \bibinfo {author} {\bibfnamefont {A.-L.}\ \bibnamefont {Barab{\'a}si}},\ and\
  \bibinfo {author} {\bibfnamefont {J.}~\bibnamefont {Kert{\'e}sz}},\
  }\bibfield  {title} {\bibinfo {title} {Universal features of correlated
  bursty behaviour},\ }\href {https://doi.org/10.1038/srep00397} {\bibfield
  {journal} {\bibinfo  {journal} {Sci. Rep.}\ }\textbf {\bibinfo {volume}
  {2}},\ \bibinfo {pages} {397} (\bibinfo {year} {2012})}\BibitemShut {NoStop}%
\bibitem [{\citenamefont {Holme}\ and\ \citenamefont
  {Saram{\"a}ki}(2012)}]{holme2012temporal}%
  \BibitemOpen
  \bibfield  {author} {\bibinfo {author} {\bibfnamefont {P.}~\bibnamefont
  {Holme}}\ and\ \bibinfo {author} {\bibfnamefont {J.}~\bibnamefont
  {Saram{\"a}ki}},\ }\bibfield  {title} {\bibinfo {title} {Temporal networks},\
  }\href {https://doi.org/10.1016/j.physrep.2012.03.001} {\bibfield  {journal}
  {\bibinfo  {journal} {Phys. Rep.}\ }\textbf {\bibinfo {volume} {519}},\
  \bibinfo {pages} {97} (\bibinfo {year} {2012})}\BibitemShut {NoStop}%
\bibitem [{\citenamefont {Cattuto}\ \emph {et~al.}(2010)\citenamefont
  {Cattuto}, \citenamefont {{Van den Broeck}}, \citenamefont {Barrat},
  \citenamefont {Colizza}, \citenamefont {Pinton},\ and\ \citenamefont
  {Vespignani}}]{cattuto2010dynamics}%
  \BibitemOpen
  \bibfield  {author} {\bibinfo {author} {\bibfnamefont {C.}~\bibnamefont
  {Cattuto}}, \bibinfo {author} {\bibfnamefont {W.}~\bibnamefont {{Van den
  Broeck}}}, \bibinfo {author} {\bibfnamefont {A.}~\bibnamefont {Barrat}},
  \bibinfo {author} {\bibfnamefont {V.}~\bibnamefont {Colizza}}, \bibinfo
  {author} {\bibfnamefont {J.-F.}\ \bibnamefont {Pinton}},\ and\ \bibinfo
  {author} {\bibfnamefont {A.}~\bibnamefont {Vespignani}},\ }\bibfield  {title}
  {\bibinfo {title} {Dynamics of person-to-person interactions from distributed
  rfid sensor networks},\ }\href {https://doi.org/10.1371/journal.pone.0011596}
  {\bibfield  {journal} {\bibinfo  {journal} {PLOS ONE}\ }\textbf {\bibinfo
  {volume} {5}},\ \bibinfo {pages} {e11596} (\bibinfo {year}
  {2010})}\BibitemShut {NoStop}%
\bibitem [{\citenamefont {Stehl{\'e}}\ \emph {et~al.}(2010)\citenamefont
  {Stehl{\'e}}, \citenamefont {Barrat},\ and\ \citenamefont
  {Bianconi}}]{stehle2010dynamical}%
  \BibitemOpen
  \bibfield  {author} {\bibinfo {author} {\bibfnamefont {J.}~\bibnamefont
  {Stehl{\'e}}}, \bibinfo {author} {\bibfnamefont {A.}~\bibnamefont {Barrat}},\
  and\ \bibinfo {author} {\bibfnamefont {G.}~\bibnamefont {Bianconi}},\
  }\bibfield  {title} {\bibinfo {title} {Dynamical and bursty interactions in
  social networks},\ }\href {https://doi.org/10.1103/PhysRevE.81.035101}
  {\bibfield  {journal} {\bibinfo  {journal} {Phys. Rev. E}\ }\textbf {\bibinfo
  {volume} {81}},\ \bibinfo {pages} {035101(R)} (\bibinfo {year}
  {2010})}\BibitemShut {NoStop}%
\bibitem [{\citenamefont {Zhao}\ \emph {et~al.}(2011)\citenamefont {Zhao},
  \citenamefont {Stehl{\'e}}, \citenamefont {Bianconi},\ and\ \citenamefont
  {Barrat}}]{zhao2011social}%
  \BibitemOpen
  \bibfield  {author} {\bibinfo {author} {\bibfnamefont {K.}~\bibnamefont
  {Zhao}}, \bibinfo {author} {\bibfnamefont {J.}~\bibnamefont {Stehl{\'e}}},
  \bibinfo {author} {\bibfnamefont {G.}~\bibnamefont {Bianconi}},\ and\
  \bibinfo {author} {\bibfnamefont {A.}~\bibnamefont {Barrat}},\ }\bibfield
  {title} {\bibinfo {title} {Social network dynamics of face-to-face
  interactions},\ }\href {https://doi.org/10.1103/PhysRevE.83.056109}
  {\bibfield  {journal} {\bibinfo  {journal} {Phys. Rev. E}\ }\textbf {\bibinfo
  {volume} {83}},\ \bibinfo {pages} {056109} (\bibinfo {year}
  {2011})}\BibitemShut {NoStop}%
\bibitem [{\citenamefont {Cencetti}\ \emph {et~al.}(2020)\citenamefont
  {Cencetti}, \citenamefont {Battiston}, \citenamefont {Lepri},\ and\
  \citenamefont {Karsai}}]{cencetti2020temporala}%
  \BibitemOpen
  \bibfield  {author} {\bibinfo {author} {\bibfnamefont {G.}~\bibnamefont
  {Cencetti}}, \bibinfo {author} {\bibfnamefont {F.}~\bibnamefont {Battiston}},
  \bibinfo {author} {\bibfnamefont {B.}~\bibnamefont {Lepri}},\ and\ \bibinfo
  {author} {\bibfnamefont {M.}~\bibnamefont {Karsai}},\ }\bibfield  {title}
  {\bibinfo {title} {Temporal properties of higher-order interactions in social
  networks},\ }\href {http://arxiv.org/abs/2010.03404} {\bibfield  {journal}
  {\bibinfo  {journal} {arXiv}\ ,\ \bibinfo {pages} {2010.03404}} (\bibinfo
  {year} {2020})}\BibitemShut {NoStop}%
\bibitem [{\citenamefont {Berge}(1989)}]{berge1989hypergraphs}%
  \BibitemOpen
  \bibfield  {author} {\bibinfo {author} {\bibfnamefont {C.}~\bibnamefont
  {Berge}},\ }\href
  {https://www.elsevier.com/books/hypergraphs/berge/978-0-444-87489-4} {\emph
  {\bibinfo {title} {Hypergraphs: Combinatorics of Finite Sets}}}\ (\bibinfo
  {publisher} {{North Holland}},\ \bibinfo {year} {1989})\BibitemShut {NoStop}%
\bibitem [{\citenamefont {Battiston}\ \emph {et~al.}(2020)\citenamefont
  {Battiston}, \citenamefont {Cencetti}, \citenamefont {Iacopini},
  \citenamefont {Latora}, \citenamefont {Lucas}, \citenamefont {Patania},
  \citenamefont {Young},\ and\ \citenamefont {Petri}}]{battiston2020networks}%
  \BibitemOpen
  \bibfield  {author} {\bibinfo {author} {\bibfnamefont {F.}~\bibnamefont
  {Battiston}}, \bibinfo {author} {\bibfnamefont {G.}~\bibnamefont {Cencetti}},
  \bibinfo {author} {\bibfnamefont {I.}~\bibnamefont {Iacopini}}, \bibinfo
  {author} {\bibfnamefont {V.}~\bibnamefont {Latora}}, \bibinfo {author}
  {\bibfnamefont {M.}~\bibnamefont {Lucas}}, \bibinfo {author} {\bibfnamefont
  {A.}~\bibnamefont {Patania}}, \bibinfo {author} {\bibfnamefont {J.-G.}\
  \bibnamefont {Young}},\ and\ \bibinfo {author} {\bibfnamefont
  {G.}~\bibnamefont {Petri}},\ }\bibfield  {title} {\bibinfo {title} {Networks
  beyond pairwise interactions: {{Structure}} and dynamics},\ }\href
  {https://doi.org/10.1016/j.physrep.2020.05.004} {\bibfield  {journal}
  {\bibinfo  {journal} {Phys. Rep.}\ }\textbf {\bibinfo {volume} {874}},\
  \bibinfo {pages} {1} (\bibinfo {year} {2020})}\BibitemShut {NoStop}%
\bibitem [{\citenamefont {Torres}\ \emph {et~al.}(2020)\citenamefont {Torres},
  \citenamefont {Blevins}, \citenamefont {Bassett},\ and\ \citenamefont
  {{Eliassi-Rad}}}]{torres2020why}%
  \BibitemOpen
  \bibfield  {author} {\bibinfo {author} {\bibfnamefont {L.}~\bibnamefont
  {Torres}}, \bibinfo {author} {\bibfnamefont {A.~S.}\ \bibnamefont {Blevins}},
  \bibinfo {author} {\bibfnamefont {D.~S.}\ \bibnamefont {Bassett}},\ and\
  \bibinfo {author} {\bibfnamefont {T.}~\bibnamefont {{Eliassi-Rad}}},\
  }\bibfield  {title} {\bibinfo {title} {The why, how, and when of
  representations for complex systems},\ }\href
  {http://arxiv.org/abs/2006.02870} {\bibfield  {journal} {\bibinfo  {journal}
  {arXiv}\ ,\ \bibinfo {pages} {2006.02870}} (\bibinfo {year}
  {2020})}\BibitemShut {NoStop}%
\bibitem [{\citenamefont {Granovetter}(1978)}]{granovetter1978threshold}%
  \BibitemOpen
  \bibfield  {author} {\bibinfo {author} {\bibfnamefont {M.}~\bibnamefont
  {Granovetter}},\ }\bibfield  {title} {\bibinfo {title} {Threshold {{Models}}
  of {{Collective Behavior}}},\ }\href {https://doi.org/10.1086/226707}
  {\bibfield  {journal} {\bibinfo  {journal} {Am. J. Sociol.}\ }\textbf
  {\bibinfo {volume} {83}},\ \bibinfo {pages} {1420} (\bibinfo {year}
  {1978})}\BibitemShut {NoStop}%
\bibitem [{\citenamefont {Watts}(2002)}]{watts2002simple}%
  \BibitemOpen
  \bibfield  {author} {\bibinfo {author} {\bibfnamefont {D.~J.}\ \bibnamefont
  {Watts}},\ }\bibfield  {title} {\bibinfo {title} {A simple model of global
  cascades on random networks},\ }\href
  {https://doi.org/10.1073/pnas.082090499} {\bibfield  {journal} {\bibinfo
  {journal} {Proc. Natl. Acad. Sci. U.S.A.}\ }\textbf {\bibinfo {volume}
  {99}},\ \bibinfo {pages} {5766} (\bibinfo {year} {2002})}\BibitemShut
  {NoStop}%
\bibitem [{\citenamefont {Dodds}\ and\ \citenamefont
  {Watts}(2004)}]{dodds2004universal}%
  \BibitemOpen
  \bibfield  {author} {\bibinfo {author} {\bibfnamefont {P.~S.}\ \bibnamefont
  {Dodds}}\ and\ \bibinfo {author} {\bibfnamefont {D.~J.}\ \bibnamefont
  {Watts}},\ }\bibfield  {title} {\bibinfo {title} {Universal {{Behavior}} in a
  {{Generalized Model}} of {{Contagion}}},\ }\href
  {https://doi.org/10.1103/PhysRevLett.92.218701} {\bibfield  {journal}
  {\bibinfo  {journal} {Phys. Rev. Lett.}\ }\textbf {\bibinfo {volume} {92}},\
  \bibinfo {pages} {218701} (\bibinfo {year} {2004})}\BibitemShut {NoStop}%
\bibitem [{\citenamefont {Anttila}\ \emph {et~al.}(2017)\citenamefont
  {Anttila}, \citenamefont {Mikonranta}, \citenamefont {Ketola}, \citenamefont
  {Kaitala}, \citenamefont {Laakso},\ and\ \citenamefont
  {Ruokolainen}}]{anttila2017mechanistic}%
  \BibitemOpen
  \bibfield  {author} {\bibinfo {author} {\bibfnamefont {J.}~\bibnamefont
  {Anttila}}, \bibinfo {author} {\bibfnamefont {L.}~\bibnamefont {Mikonranta}},
  \bibinfo {author} {\bibfnamefont {T.}~\bibnamefont {Ketola}}, \bibinfo
  {author} {\bibfnamefont {V.}~\bibnamefont {Kaitala}}, \bibinfo {author}
  {\bibfnamefont {J.}~\bibnamefont {Laakso}},\ and\ \bibinfo {author}
  {\bibfnamefont {L.}~\bibnamefont {Ruokolainen}},\ }\bibfield  {title}
  {\bibinfo {title} {A mechanistic underpinning for sigmoid dose-dependent
  infection},\ }\href {https://doi.org/10.1111/oik.03242} {\bibfield  {journal}
  {\bibinfo  {journal} {Oikos}\ }\textbf {\bibinfo {volume} {126}},\ \bibinfo
  {pages} {910} (\bibinfo {year} {2017})}\BibitemShut {NoStop}%
\bibitem [{\citenamefont {Landry}\ and\ \citenamefont
  {Restrepo}(2020)}]{landry2020effect}%
  \BibitemOpen
  \bibfield  {author} {\bibinfo {author} {\bibfnamefont {N.~W.}\ \bibnamefont
  {Landry}}\ and\ \bibinfo {author} {\bibfnamefont {J.~G.}\ \bibnamefont
  {Restrepo}},\ }\bibfield  {title} {\bibinfo {title} {The effect of
  heterogeneity on hypergraph contagion models},\ }\href
  {https://doi.org/10.1063/5.0020034} {\bibfield  {journal} {\bibinfo
  {journal} {Chaos}\ }\textbf {\bibinfo {volume} {30}},\ \bibinfo {pages}
  {103117} (\bibinfo {year} {2020})}\BibitemShut {NoStop}%
\bibitem [{\citenamefont {Liu}\ \emph {et~al.}(1987)\citenamefont {Liu},
  \citenamefont {Hethcote},\ and\ \citenamefont {Levin}}]{Liu1987}%
  \BibitemOpen
  \bibfield  {author} {\bibinfo {author} {\bibfnamefont {W.-m.}\ \bibnamefont
  {Liu}}, \bibinfo {author} {\bibfnamefont {H.~W.}\ \bibnamefont {Hethcote}},\
  and\ \bibinfo {author} {\bibfnamefont {S.~A.}\ \bibnamefont {Levin}},\
  }\bibfield  {title} {\bibinfo {title} {{Dynamical behavior of epidemiological
  models with nonlinear incidence rates}},\ }\href
  {https://doi.org/10.1007/BF00277162} {\bibfield  {journal} {\bibinfo
  {journal} {J. Math. Biol.}\ }\textbf {\bibinfo {volume} {25}},\ \bibinfo
  {pages} {359} (\bibinfo {year} {1987})}\BibitemShut {NoStop}%
\bibitem [{\citenamefont {Scarpino}\ \emph {et~al.}(2016)\citenamefont
  {Scarpino}, \citenamefont {Allard},\ and\ \citenamefont
  {H{\'e}bert-Dufresne}}]{scarpino2016effect}%
  \BibitemOpen
  \bibfield  {author} {\bibinfo {author} {\bibfnamefont {S.~V.}\ \bibnamefont
  {Scarpino}}, \bibinfo {author} {\bibfnamefont {A.}~\bibnamefont {Allard}},\
  and\ \bibinfo {author} {\bibfnamefont {L.}~\bibnamefont
  {H{\'e}bert-Dufresne}},\ }\bibfield  {title} {\bibinfo {title} {The effect of
  a prudent adaptive behaviour on disease transmission},\ }\href
  {https://doi.org/10.1038/nphys3832} {\bibfield  {journal} {\bibinfo
  {journal} {Nat. Phys.}\ }\textbf {\bibinfo {volume} {12}},\ \bibinfo {pages}
  {1042} (\bibinfo {year} {2016})}\BibitemShut {NoStop}%
\bibitem [{\citenamefont {Iacopini}\ \emph {et~al.}(2019)\citenamefont
  {Iacopini}, \citenamefont {Petri}, \citenamefont {Barrat},\ and\
  \citenamefont {Latora}}]{Iacopini2019}%
  \BibitemOpen
  \bibfield  {author} {\bibinfo {author} {\bibfnamefont {I.}~\bibnamefont
  {Iacopini}}, \bibinfo {author} {\bibfnamefont {G.}~\bibnamefont {Petri}},
  \bibinfo {author} {\bibfnamefont {A.}~\bibnamefont {Barrat}},\ and\ \bibinfo
  {author} {\bibfnamefont {V.}~\bibnamefont {Latora}},\ }\bibfield  {title}
  {\bibinfo {title} {Simplicial models of social contagion},\ }\href
  {https://doi.org/10.1038/s41467-019-10431-6} {\bibfield  {journal} {\bibinfo
  {journal} {Nat. Commun.}\ }\textbf {\bibinfo {volume} {10}},\ \bibinfo
  {pages} {2485} (\bibinfo {year} {2019})}\BibitemShut {NoStop}%
\bibitem [{\citenamefont {Jhun}\ \emph {et~al.}(2019)\citenamefont {Jhun},
  \citenamefont {Jo},\ and\ \citenamefont {Kahng}}]{jhun2019simplicial}%
  \BibitemOpen
  \bibfield  {author} {\bibinfo {author} {\bibfnamefont {B.}~\bibnamefont
  {Jhun}}, \bibinfo {author} {\bibfnamefont {M.}~\bibnamefont {Jo}},\ and\
  \bibinfo {author} {\bibfnamefont {B.}~\bibnamefont {Kahng}},\ }\bibfield
  {title} {\bibinfo {title} {Simplicial {{SIS}} model in scale-free uniform
  hypergraph},\ }\href {https://doi.org/10.1088/1742-5468/ab5367} {\bibfield
  {journal} {\bibinfo  {journal} {J. Stat. Mech.}\ }\textbf {\bibinfo {volume}
  {2019}},\ \bibinfo {pages} {123207} (\bibinfo {year} {2019})}\BibitemShut
  {NoStop}%
\bibitem [{\citenamefont {{de Arruda}}\ \emph {et~al.}(2020)\citenamefont {{de
  Arruda}}, \citenamefont {Petri},\ and\ \citenamefont
  {Moreno}}]{ferrazdearruda2020social}%
  \BibitemOpen
  \bibfield  {author} {\bibinfo {author} {\bibfnamefont {G.~F.}\ \bibnamefont
  {{de Arruda}}}, \bibinfo {author} {\bibfnamefont {G.}~\bibnamefont {Petri}},\
  and\ \bibinfo {author} {\bibfnamefont {Y.}~\bibnamefont {Moreno}},\
  }\bibfield  {title} {\bibinfo {title} {Social contagion models on
  hypergraphs},\ }\href {https://doi.org/10.1103/PhysRevResearch.2.023032}
  {\bibfield  {journal} {\bibinfo  {journal} {Phys. Rev. Research}\ }\textbf
  {\bibinfo {volume} {2}},\ \bibinfo {pages} {023032} (\bibinfo {year}
  {2020})}\BibitemShut {NoStop}%
\bibitem [{\citenamefont {H\'ebert-Dufresne}\ \emph {et~al.}(2010)\citenamefont
  {H\'ebert-Dufresne}, \citenamefont {No\"el}, \citenamefont {Marceau},
  \citenamefont {Allard},\ and\ \citenamefont
  {Dub\'e}}]{hebert2010propagation}%
  \BibitemOpen
  \bibfield  {author} {\bibinfo {author} {\bibfnamefont {L.}~\bibnamefont
  {H\'ebert-Dufresne}}, \bibinfo {author} {\bibfnamefont {P.-A.}\ \bibnamefont
  {No\"el}}, \bibinfo {author} {\bibfnamefont {V.}~\bibnamefont {Marceau}},
  \bibinfo {author} {\bibfnamefont {A.}~\bibnamefont {Allard}},\ and\ \bibinfo
  {author} {\bibfnamefont {L.~J.}\ \bibnamefont {Dub\'e}},\ }\bibfield  {title}
  {\bibinfo {title} {Propagation dynamics on networks featuring complex
  topologies},\ }\href {https://doi.org/10.1103/PhysRevE.82.036115} {\bibfield
  {journal} {\bibinfo  {journal} {Phys. Rev. E}\ }\textbf {\bibinfo {volume}
  {82}},\ \bibinfo {pages} {036115} (\bibinfo {year} {2010})}\BibitemShut
  {NoStop}%
\bibitem [{\citenamefont {Marceau}\ \emph {et~al.}(2010)\citenamefont
  {Marceau}, \citenamefont {No\"el}, \citenamefont {H\'ebert-Dufresne},
  \citenamefont {Allard},\ and\ \citenamefont {Dub\'e}}]{marceau2010adaptive}%
  \BibitemOpen
  \bibfield  {author} {\bibinfo {author} {\bibfnamefont {V.}~\bibnamefont
  {Marceau}}, \bibinfo {author} {\bibfnamefont {P.-A.}\ \bibnamefont {No\"el}},
  \bibinfo {author} {\bibfnamefont {L.}~\bibnamefont {H\'ebert-Dufresne}},
  \bibinfo {author} {\bibfnamefont {A.}~\bibnamefont {Allard}},\ and\ \bibinfo
  {author} {\bibfnamefont {L.~J.}\ \bibnamefont {Dub\'e}},\ }\bibfield  {title}
  {\bibinfo {title} {Adaptive networks: Coevolution of disease and topology},\
  }\href {https://doi.org/10.1103/PhysRevE.82.036116} {\bibfield  {journal}
  {\bibinfo  {journal} {Phys. Rev. E}\ }\textbf {\bibinfo {volume} {82}},\
  \bibinfo {pages} {036116} (\bibinfo {year} {2010})}\BibitemShut {NoStop}%
\bibitem [{\citenamefont {Gleeson}(2011)}]{Gleeson2011}%
  \BibitemOpen
  \bibfield  {author} {\bibinfo {author} {\bibfnamefont {J.~P.}\ \bibnamefont
  {Gleeson}},\ }\bibfield  {title} {\bibinfo {title} {High-accuracy
  approximation of binary-state dynamics on networks},\ }\href
  {https://doi.org/10.1103/PhysRevLett.107.068701} {\bibfield  {journal}
  {\bibinfo  {journal} {Phys. Rev. Lett.}\ }\textbf {\bibinfo {volume} {107}},\
  \bibinfo {pages} {068701} (\bibinfo {year} {2011})}\BibitemShut {NoStop}%
\bibitem [{\citenamefont {Lindquist}\ \emph {et~al.}(2011)\citenamefont
  {Lindquist}, \citenamefont {Ma}, \citenamefont {van~den Driessche},\ and\
  \citenamefont {Willeboordse}}]{Lindquist2011}%
  \BibitemOpen
  \bibfield  {author} {\bibinfo {author} {\bibfnamefont {J.}~\bibnamefont
  {Lindquist}}, \bibinfo {author} {\bibfnamefont {J.}~\bibnamefont {Ma}},
  \bibinfo {author} {\bibfnamefont {P.}~\bibnamefont {van~den Driessche}},\
  and\ \bibinfo {author} {\bibfnamefont {F.~H.}\ \bibnamefont {Willeboordse}},\
  }\bibfield  {title} {\bibinfo {title} {{Effective degree network disease
  models}},\ }\href {https://doi.org/10.1007/s00285-010-0331-2} {\bibfield
  {journal} {\bibinfo  {journal} {J. Math. Biol.}\ }\textbf {\bibinfo {volume}
  {62}},\ \bibinfo {pages} {143} (\bibinfo {year} {2011})}\BibitemShut
  {NoStop}%
\bibitem [{\citenamefont {O'Sullivan}\ \emph {et~al.}(2015)\citenamefont
  {O'Sullivan}, \citenamefont {O'Keeffe}, \citenamefont {Fennell},\ and\
  \citenamefont {Gleeson}}]{osullivan2015mathematical}%
  \BibitemOpen
  \bibfield  {author} {\bibinfo {author} {\bibfnamefont {D.~J.~P.}\
  \bibnamefont {O'Sullivan}}, \bibinfo {author} {\bibfnamefont {G.~J.}\
  \bibnamefont {O'Keeffe}}, \bibinfo {author} {\bibfnamefont {P.~G.}\
  \bibnamefont {Fennell}},\ and\ \bibinfo {author} {\bibfnamefont {J.~P.}\
  \bibnamefont {Gleeson}},\ }\bibfield  {title} {\bibinfo {title} {Mathematical
  modeling of complex contagion on clustered networks},\ }\href
  {https://doi.org/10.3389/fphy.2015.00071} {\bibfield  {journal} {\bibinfo
  {journal} {Front. Phys.}\ }\textbf {\bibinfo {volume} {3}},\ \bibinfo {pages}
  {71} (\bibinfo {year} {2015})}\BibitemShut {NoStop}%
\bibitem [{\citenamefont {St-Onge}\ \emph
  {et~al.}(2021{\natexlab{a}})\citenamefont {St-Onge}, \citenamefont
  {Thibeault}, \citenamefont {Allard}, \citenamefont {Dub{\'e}},\ and\
  \citenamefont {H{\'e}bert-Dufresne}}]{st-onge2021social}%
  \BibitemOpen
  \bibfield  {author} {\bibinfo {author} {\bibfnamefont {G.}~\bibnamefont
  {St-Onge}}, \bibinfo {author} {\bibfnamefont {V.}~\bibnamefont {Thibeault}},
  \bibinfo {author} {\bibfnamefont {A.}~\bibnamefont {Allard}}, \bibinfo
  {author} {\bibfnamefont {L.~J.}\ \bibnamefont {Dub{\'e}}},\ and\ \bibinfo
  {author} {\bibfnamefont {L.}~\bibnamefont {H{\'e}bert-Dufresne}},\ }\bibfield
   {title} {\bibinfo {title} {Social confinement and mesoscopic localization of
  epidemics on networks},\ }\href
  {https://doi.org/10.1103/PhysRevLett.126.098301} {\bibfield  {journal}
  {\bibinfo  {journal} {Phys. Rev. Lett.}\ }\textbf {\bibinfo {volume} {126}},\
  \bibinfo {pages} {098301} (\bibinfo {year} {2021}{\natexlab{a}})}\BibitemShut
  {NoStop}%
\bibitem [{\citenamefont {St-Onge}\ \emph
  {et~al.}(2021{\natexlab{b}})\citenamefont {St-Onge}, \citenamefont
  {Thibeault}, \citenamefont {Allard}, \citenamefont {Dub{\'e}},\ and\
  \citenamefont {H{\'e}bert-Dufresne}}]{st-onge2021master}%
  \BibitemOpen
  \bibfield  {author} {\bibinfo {author} {\bibfnamefont {G.}~\bibnamefont
  {St-Onge}}, \bibinfo {author} {\bibfnamefont {V.}~\bibnamefont {Thibeault}},
  \bibinfo {author} {\bibfnamefont {A.}~\bibnamefont {Allard}}, \bibinfo
  {author} {\bibfnamefont {L.~J.}\ \bibnamefont {Dub{\'e}}},\ and\ \bibinfo
  {author} {\bibfnamefont {L.}~\bibnamefont {H{\'e}bert-Dufresne}},\ }\bibfield
   {title} {\bibinfo {title} {Master equation analysis of mesoscopic
  localization in contagion dynamics on higher-order networks},\ }\href
  {https://doi.org/10.1103/PhysRevE.103.032301} {\bibfield  {journal} {\bibinfo
   {journal} {Phys. Rev. E}\ }\textbf {\bibinfo {volume} {103}},\ \bibinfo
  {pages} {032301} (\bibinfo {year} {2021}{\natexlab{b}})}\BibitemShut
  {NoStop}%
\bibitem [{\citenamefont {Nadarajah}(2011)}]{nadarajah2011exact}%
  \BibitemOpen
  \bibfield  {author} {\bibinfo {author} {\bibfnamefont {S.}~\bibnamefont
  {Nadarajah}},\ }\bibfield  {title} {\bibinfo {title} {Exact distribution of
  the product of m gamma and n {{Pareto}} random variables},\ }\href
  {https://doi.org/10.1016/j.cam.2011.04.018} {\bibfield  {journal} {\bibinfo
  {journal} {J. Comput. Appl. Math.}\ }\textbf {\bibinfo {volume} {235}},\
  \bibinfo {pages} {4496} (\bibinfo {year} {2011})}\BibitemShut {NoStop}%
\bibitem [{\citenamefont {Fosdick}\ \emph {et~al.}(2018)\citenamefont
  {Fosdick}, \citenamefont {Larremore}, \citenamefont {Nishimura},\ and\
  \citenamefont {Ugander}}]{fosdick2018configuring}%
  \BibitemOpen
  \bibfield  {author} {\bibinfo {author} {\bibfnamefont {B.~K.}\ \bibnamefont
  {Fosdick}}, \bibinfo {author} {\bibfnamefont {D.~B.}\ \bibnamefont
  {Larremore}}, \bibinfo {author} {\bibfnamefont {J.}~\bibnamefont
  {Nishimura}},\ and\ \bibinfo {author} {\bibfnamefont {J.}~\bibnamefont
  {Ugander}},\ }\bibfield  {title} {\bibinfo {title} {Configuring random graph
  models with fixed degree sequences},\ }\href
  {https://doi.org/10.1137/16M1087175} {\bibfield  {journal} {\bibinfo
  {journal} {SIAM Rev.}\ }\textbf {\bibinfo {volume} {60}},\ \bibinfo {pages}
  {315} (\bibinfo {year} {2018})}\BibitemShut {NoStop}%
\bibitem [{\citenamefont {Chodrow}(2020)}]{chodrow2020configuration}%
  \BibitemOpen
  \bibfield  {author} {\bibinfo {author} {\bibfnamefont {P.~S.}\ \bibnamefont
  {Chodrow}},\ }\bibfield  {title} {\bibinfo {title} {Configuration models of
  random hypergraphs},\ }\href {https://doi.org/10.1093/comnet/cnaa018}
  {\bibfield  {journal} {\bibinfo  {journal} {J. Complex Netw.}\ }\textbf
  {\bibinfo {volume} {8}},\ \bibinfo {pages} {cnaa018} (\bibinfo {year}
  {2020})}\BibitemShut {NoStop}%
\bibitem [{\citenamefont {de~Oliveira}\ and\ \citenamefont
  {Dickman}(2005)}]{deoliveira2005}%
  \BibitemOpen
  \bibfield  {author} {\bibinfo {author} {\bibfnamefont {M.~M.}\ \bibnamefont
  {de~Oliveira}}\ and\ \bibinfo {author} {\bibfnamefont {R.}~\bibnamefont
  {Dickman}},\ }\bibfield  {title} {\bibinfo {title} {How to simulate the
  quasistationary state},\ }\href {https://doi.org/10.1103/PhysRevE.71.016129}
  {\bibfield  {journal} {\bibinfo  {journal} {Phys. Rev. E}\ }\textbf {\bibinfo
  {volume} {71}},\ \bibinfo {pages} {016129} (\bibinfo {year}
  {2005})}\BibitemShut {NoStop}%
\bibitem [{\citenamefont {St-Onge}\ \emph
  {et~al.}(2021{\natexlab{c}})\citenamefont {St-Onge}, \citenamefont {Sun},
  \citenamefont {Allard}, \citenamefont {H{\'{e}}bert-Dufresne},\ and\
  \citenamefont {Bianconi}}]{SSdata_curated}%
  \BibitemOpen
  \bibfield  {author} {\bibinfo {author} {\bibfnamefont {G.}~\bibnamefont
  {St-Onge}}, \bibinfo {author} {\bibfnamefont {H.}~\bibnamefont {Sun}},
  \bibinfo {author} {\bibfnamefont {A.}~\bibnamefont {Allard}}, \bibinfo
  {author} {\bibfnamefont {L.}~\bibnamefont {H{\'{e}}bert-Dufresne}},\ and\
  \bibinfo {author} {\bibfnamefont {G.}~\bibnamefont {Bianconi}},\ }\href
  {https://docs.google.com/spreadsheets/d/1TTX6bOOCIDA9StfHcP8Kn809551cbBjn1rCVzg3TCzM/edit?usp=sharing}
  {\bibinfo {title} {Superspreading events data from arxiv:2101.07229}}
  (\bibinfo {year} {2021}{\natexlab{c}}),\ \bibinfo {note} {updated:
  2021-02-01}\BibitemShut {NoStop}%
\bibitem [{\citenamefont {St-Onge}\ \emph
  {et~al.}(2021{\natexlab{d}})\citenamefont {St-Onge}, \citenamefont {Sun},
  \citenamefont {Allard}, \citenamefont {H{\'{e}}bert-Dufresne},\ and\
  \citenamefont {Bianconi}}]{github-repository}%
  \BibitemOpen
  \bibfield  {author} {\bibinfo {author} {\bibfnamefont {G.}~\bibnamefont
  {St-Onge}}, \bibinfo {author} {\bibfnamefont {H.}~\bibnamefont {Sun}},
  \bibinfo {author} {\bibfnamefont {A.}~\bibnamefont {Allard}}, \bibinfo
  {author} {\bibfnamefont {L.}~\bibnamefont {H{\'{e}}bert-Dufresne}},\ and\
  \bibinfo {author} {\bibfnamefont {G.}~\bibnamefont {Bianconi}},\ }\href
  {https://github.com/gstonge/heterogeneous-exposure-hons} {\bibinfo {title}
  {heterogeneous-exposure-hons}} (\bibinfo {year}
  {2021}{\natexlab{d}})\BibitemShut {NoStop}%
\end{thebibliography}
\end{document}